\documentclass[acmtog, nonacm]{acmart}

\title{Monte Carlo Rendering to Diffusion Curves with Differential BEM}

\author{Ryusuke Sugimoto}
\orcid{0000-0001-5894-0423}
\affiliation{
    \institution{University of Waterloo}
    \city{Waterloo}
    \country{Canada}
}
\email{rsugimot@uwaterloo.ca}
\author{Christopher Batty}
\orcid{0000-0003-3830-7772}
\affiliation{
    \institution{University of Waterloo}
    \city{Waterloo}
    \country{Canada}
}
\email{c2batty@uwaterloo.ca}

\author{Siddhartha Chaudhuri}
\orcid{0009-0009-8588-1436}
\affiliation{
    \institution{Adobe Research}
    \city{New York}
    \country{United States of America}
}
\email{sidch@adobe.com}

\author{Iliyan Georgiev}
\orcid{0000-0002-9655-2138}
\affiliation{
    \institution{Adobe Research}
    \city{London}
    \country{United Kingdom}
}
\email{igeorgiev@adobe.com}

\author{Toshiya Hachisuka}
\orcid{0000-0003-0284-3776}
\affiliation{
    \institution{University of Waterloo}
    \city{Waterloo}
    \country{Canada}
}
\email{thachisu@uwaterloo.ca}

\author{Kevin Wampler}
\orcid{0009-0005-1780-6814}
\affiliation{
    \institution{Adobe Research}
    \city{Seattle}
    \country{United States of America}
}
\email{kwampler@adobe.com}

\author{Michal Lukáč}
\orcid{0000-0002-9664-7786}
\affiliation{
    \institution{Adobe Research}
    \city{San Jose}
    \country{United States of America}
}
\email{mike.k.lukac@gmail.com}



\usepackage{wrapfig}
\usepackage{hyperref}
\usepackage{cleveref}
\usepackage{svg}
\usepackage[utf8]{inputenc}
\usepackage{mathtools}
\usepackage{subcaption}
\usepackage{tikz}
\usepackage{pgfplots}
\usetikzlibrary{pgfplots.groupplots}

\newcommand{\parm}{\theta}
\newcommand{\vecx}{\mathbf{x}}
\newcommand{\vecy}{\mathbf{y}}
\newcommand{\vecz}{\mathbf{z}}

\newcommand{\vecn}{\mathbf{n}}
\newcommand{\dx}{\,\mathrm{d}\vecx}
\newcommand{\dy}{\,\mathrm{d}\vecy}
\newcommand{\dz}{\,\mathrm{d}\vecz}
\newcommand{\handlecolor}{\mathbf{c}}

\newcommand{\vecf}{\mathbf{f}}

\newcommand{\weight}{\mathbf{w}}
\newcommand{\loss}{\mathcal{L}}
\newcommand{\domain}{\Omega}
\newcommand{\imagedomain}{\mathcal{A}}
\newcommand{\handleset}{\mathcal{H}}
\newcommand{\handleterm}{\vecf}

\newcommand{\transpose}{^\top}
\newcommand{\sol}{\mathbf{u}}
\renewcommand{\laplace}{\Delta}

\newcommand{\damping}{\lambda}
\newcommand{\discrete}[1]{\mathsf{#1}}
\newcommand{\weightc}{\weight_\text{c}}
\newcommand{\weightd}{\weight_\text{d}}
\newcommand{\itar}{I_t}
\newcommand{\ireconst}{\sol}

\crefformat{equation}{Eq.~#2#1#3}
\crefformat{figure}{Fig.~#2#1#3}
\crefformat{section}{Sec.~#2#1#3}

\definecolor{mycolor1}{HTML}{6A99D0}
\definecolor{mycolor2}{HTML}{DE8344}
\definecolor{mycolor3}{HTML}{7EAB55}
\definecolor{mycolor4}{HTML}{F5C342}
\definecolor{mycolord1}{HTML}{4F7CB1} 
\definecolor{mycolord2}{HTML}{C36E3A} 
\definecolor{mycolord3}{HTML}{659142}
\definecolor{mycolord4}{HTML}{D4A91E}

\begin{abstract}
We present a method for generating vector graphics, in the form of diffusion curves, directly from noisy samples produced by a Monte Carlo renderer.
While generating raster images from 3D geometry via Monte Carlo raytracing is commonplace, there is no corresponding practical approach for robustly and directly extracting editable vector images with shading information from 3D geometry.
To fill this gap, we formulate the problem as a stochastic optimization problem over the space of geometries and colors of diffusion curve handles, and solve it with the Levenberg–Marquardt algorithm. 
At the core of our method is a novel differential boundary element method (BEM) framework that reconstructs colors from diffusion curve handles and computes gradients with respect to their parameters, requiring the expensive matrix factorization only once at the beginning of the optimization.
Unlike triangulation-based techniques that require a clean domain decomposition, our method is robust to geometrically challenging scenarios, such as intersecting diffusion curves, and to color noise in the target image, enabling the direct use of noisy Monte Carlo samples without requiring a converged, error-free input image.
We demonstrate the robustness and broad applicability of our approach across several test cases.
Finally, we highlight several open questions raised by our work, which spans both theory and applications.
\end{abstract}

\begin{teaserfigure}
  \centering
\includegraphics[width=0.24\linewidth]{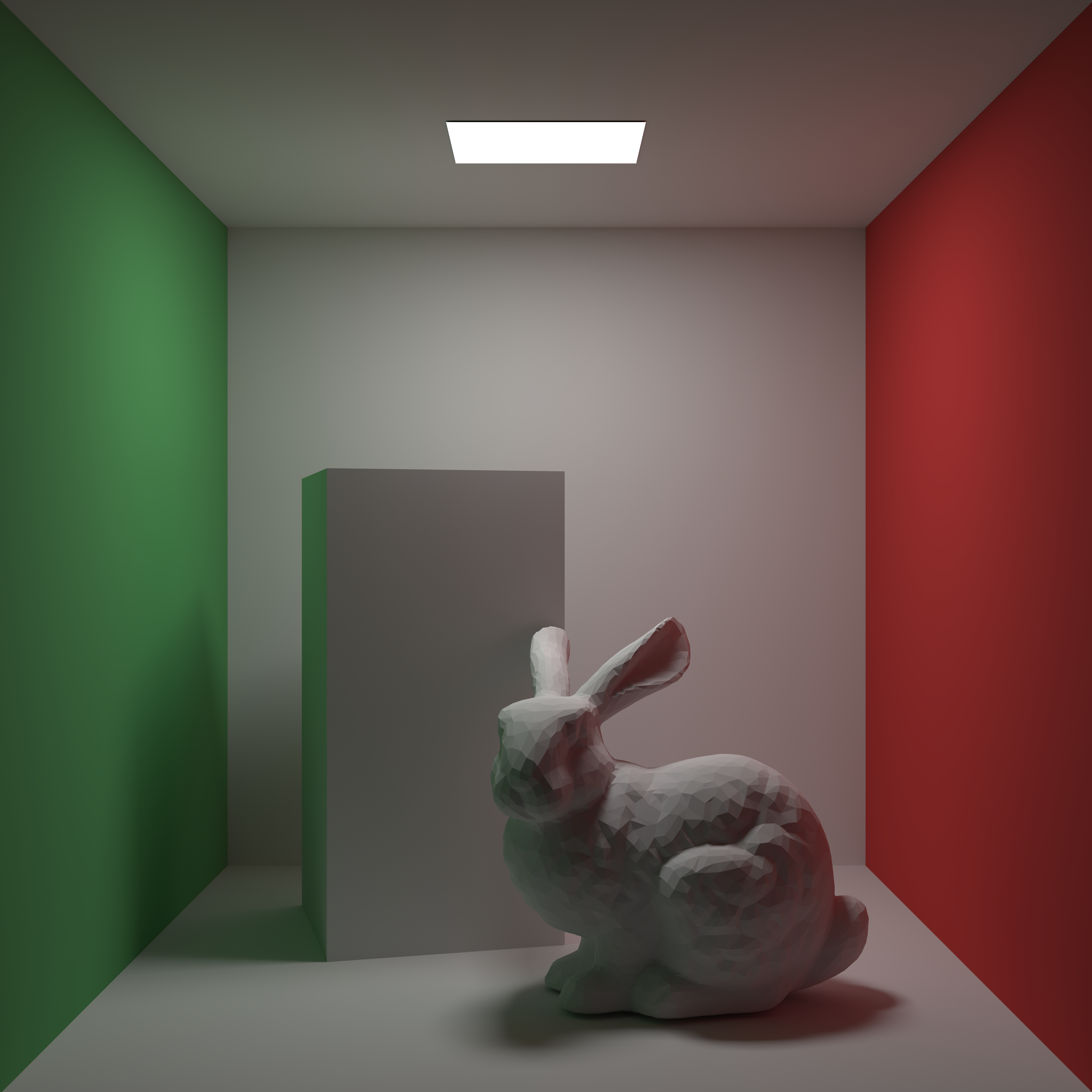}
\includegraphics[width=0.24\linewidth]{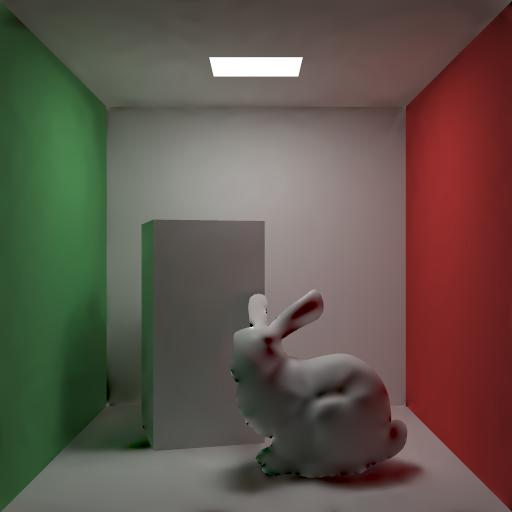}
\includesvg[width=0.24\linewidth]{Figures/results/cb2/cb2_baseline_100.svg}%
\begin{minipage}[b]{0.24\linewidth}
\centering
    \resizebox{1.0\linewidth}{!}{
    \begin{tikzpicture}
    \begin{axis}[width=\linewidth,height=1.1\linewidth,
    ymode=log, log basis y=10,
    grid=both, scale only axis,
    xtick={0, 25, 50, 75, 100},
    ytick={1, 0.31622776601, 0.1, 0.05623413251, 0.031622776601, 0.0177827941, 0.01},
    xmin=0,
    xmax=100,
    axis background/.style={fill=gray!5},
    every axis x label/.style={at={(0.5, -0.1)},anchor=north, font=\sffamily\bfseries\small},
    every axis y label/.style={at={(0, 1)},anchor=north east, font=\sffamily\bfseries\small},
    ylabel=RMSE,
    xlabel=Step
    ]
    \addplot+[line width=2pt,mark=none,color=mycolor1]table[]{Figures/results/cb2_data/baseline_step_vs_rmse_tonemapped_abs.tex};
    \end{axis}
    \end{tikzpicture}}
\end{minipage}\\
\begin{minipage}[t]{0.24\linewidth}\centering\textsf{\small (a) Target}\end{minipage}
\begin{minipage}[t]{0.24\linewidth}\centering\textsf{\small (b) Reconstructed}\end{minipage}
\begin{minipage}[t]{0.24\linewidth}\centering\textsf{\small (c) Diffusion Curves}\end{minipage}
\begin{minipage}[t]{0.24\linewidth}\centering\textsf{\small (d) Error Curve}\end{minipage}\\
  \caption{Given the ability to query Monte Carlo renderers' noisy samples of the target image (a), we directly extract a sparse set of diffusion curves (c) that reconstruct the image (b), using an optimization approach (d), which progressively reduces a reconstruction loss. }
  \label{fig:teaser}
\end{teaserfigure}

\begin{document}
\maketitle

\section{Introduction}
It is standard practice to render a 3D scene into a \emph{raster} image, and numerous rendering algorithms and sampling techniques have been developed to support this process. However, there is often a significant amount of redundancy in raster image representation, particularly when the target image is mostly smooth with sparse discontinuities (i.e., piecewise smooth). Raster images also introduce resolution-related inaccuracies due to the need for a fixed discretization scale. By contrast, a \emph{vector} image encodes visual information using geometric primitives, such as line segments, allowing for resolution-independent, compact, and intuitively editable representations. Prior work on constructing vector images from 3D scenes typically requires one to first generate an intermediate, full-resolution and (nearly) noise-free raster image; that raster image is then vectorized. 

We propose a technique for rendering a 3D scene \emph{directly} into a vector image, bypassing the rasterization stage entirely~(\cref{fig:teaser}). As in many modern rendering pipelines, we assume a Monte Carlo renderer that produces \emph{noisy color samples} from the 3D scene.
In particular, we focus on scenes with relatively simple geometry and without high-frequency detail, which are well-suited to vector formats. Even in such cases, it is increasingly common to use Monte Carlo rendering to capture global illumination effects, e.g., colored soft shadows and reflections.
Unlike conventional approaches, however, our method avoids rasterization at any stage, thereby eliminating resolution limits and robustness issues associated with vectorization of raster images.

We adopt \emph{diffusion curves} \cite{Orzan:Diffusion:2008, Jeschke:2009:GPULaplacian} as our vector image representation. A diffusion curve image is defined by a set of curves in 2D space (also referred to as \emph{handles}), where each curve encodes color information on its two sides. The final image is obtained by diffusing these colored boundary conditions throughout the domain, formally as the solution to the \emph{Laplace equation}. Our task is to solve the \emph{inverse problem}: extract the diffusion curve representation for an image given by Monte Carlo samples. We therefore adopt a gradient-based optimization approach, which demands the ability to evaluate two key quantities: (1) the reconstructed color at any point given a set of diffusion curve handles, and (2) the \emph{gradient} of that color with respect to the handles’ geometric and color parameters. 

To support both, we develop a new \emph{differential boundary element method} (differential BEM) as an extension of the traditional BEM~\cite{Liu2012:BEMSurvey, Clemens:BEM2013} that only considers the forward (Laplace) problem (i.e., estimating a solution given boundaries), and show its application to our problem setting. 
Aside from easily extracted geometric contour information, we only have access to noisy Monte Carlo color estimates from the renderer. Thus, we are not provided with an initial guess for additional handle placement.
Our gradient-based optimization approach with differential BEM supports optimization under this challenging situation by starting with many unstructured and often intersecting handles, and progressively optimizing and pruning them.
Importantly, the chosen differential BEM formulation mostly sidesteps costly dense matrix factorization by optimizing a carefully selected set of parameters, thereby eliminating the need to factorize matrices tied to the handle degrees of freedom at every iteration.
For optimization, we employ the \emph{Levenberg–Marquardt} method, which is derived from a least squares loss defined over the continuous image space. We use Monte Carlo integration to estimate the approximate Hessian and gradient required during the optimization.\looseness=-1

Together, these components form a principled and robust pipeline for  \emph{progressively} extracting vector graphics representations directly from noisy Monte Carlo samples, without relying on raster or mesh-based intermediaries.
In summary, our contributions are:
\begin{itemize}
\item A novel, efficient differential boundary element method for the Laplace equation with double-sided boundaries (\cref{sec:bem}), and
\item A rasterization-free system that directly extracts diffusion curves from Monte Carlo renderer outputs using Levenberg–Marquardt optimization (\cref{sec:optimization}).
\end{itemize}

\section{Diffusion Curves}
\label{sec:diffusioncurves}
A wide variety of vector image representations have been developed over the years; we refer readers to the survey by \citet{Tian:2024:VectorGraphicsSurvey} for a comprehensive overview. In our work, we focus specifically on the diffusion curves representation~\cite{Orzan:Diffusion:2008, Jeschke:2009:GPULaplacian}, which models piecewise smooth color transitions via solutions to elliptic PDEs. In this section, we summarize key advances in diffusion curves to provide context for our method.

\subsection{Forward Problem}
 Given a set of colored \emph{handles} (e.g., line segments, curves) placed in a domain $\domain$ within the image space, the diffusion curves approach diffuses the colors from the handles to define the reconstructed colors at any point within the domain. Formally, given a set of handles whose geometries and colors are parameterized by $\theta$, we define the reconstructed image color $\ireconst(\vecx; \parm)$ to be the solution $\sol(\vecx)$ of the Laplace equation:
\begin{equation}\label{eq:laplace}
    \laplace \sol(\vecx) = 0 \quad \text{for } \vecx \in \domain\backslash\handleset(\parm),
\end{equation}
where $\handleset(\parm)$ is the set of all handles. We use $\ireconst(\vecx; \parm)$ and $\sol(\vecx)$ interchangeably throughout this text, preferring $\ireconst(\vecx; \parm)$ when we wish to emphasize the parameter dependence of the reconstructed color.
At the handles, it is typical to specify a double-sided Dirichlet boundary condition,
\begin{equation}\label{eq:dirichlet}
    \sol(\vecx) = \{
        \handlecolor^+(\vecx;\parm), \;
        \handlecolor^-(\vecx;\parm)\}
     \quad \text{for } \vecx \in \handleset(\parm),
\end{equation}
where the colors $\handlecolor^\pm(\vecx;\parm) $ are generally two distinct colors that the solution takes on as $\vecx$ approaches the curve from one side or the other.
While our method supports double-sided boundaries, we will consider a different specification for them, as we describe in \cref{sec:bem}.
The handles may be specified as curves (e.g., piecewise linear, B\'{e}zier curves, etc.) in which case the color can be defined to take different values on either side of the curve, and to vary along the length of the curve.
For this paper, we restrict our representation to supporting only line segments for simplicity.

Additionally, image space domain boundaries are treated with the
Neumann condition,
\begin{equation}\label{eq:neumann}
    \frac{\partial \sol}{\partial \vecn}(\vecx) = \mathbf{0} \quad \text{for } \vecx \in \partial\domain,
\end{equation}
where the normal vector $\vecn$ points outwards from the domain. These boundaries act as color barriers rather than sources.

Because images reconstructed in this manner exhibit sharp color discontinuities or peaks at handle positions, \citet{Jeschke:2016:Generalized}, \citet{Orzan:Diffusion:2008}, and \citet{Zhao:2018:InverseDiffusion} additionally compute the size of a  spatially-varying (Gaussian) smoothing filter as an additional handle parameter and apply the filter to the image.

While the original work by \citet{Orzan:Diffusion:2008} and ~\citet{Jeschke:2009:GPULaplacian} employed a finite difference method, the core of the diffusion curves formulation lies in its underlying continuous mathematical formulation, which allows for various alternative solvers, including finite element methods~\cite{Boye:2012:Vectorial,Zhao:2018:InverseDiffusion}, boundary element methods~\cite{Bang:Diffusion:2023, Chen:2025:Lightning, Sun:2012:Texture, Sun:2014:FastMultipole, Gronde:DiffusionCurves:2010}, %
method of fundamental solutions~\cite{Chen:2024:Lightning},
triangle interpolation-based methods \cite{Pang2012:FastRendering}, ray tracing-based methods~\cite{Bowers:2011:RayTracing, Prevost:2015:Vectorial}, and pointwise Monte Carlo methods~\cite{Sawhney:WoS:2020, Sugimoto:WoB:2023}.
We design a method based on BEM to optimize handles, but the resulting handles can be used as input to any type of forward solver.

\subsection{Inverse Problem}

Diffusion curves can compactly represent smooth images with a few handles, provided the level of detail of the target image is not too high.
Obtaining such a sparse set of handles given a target image representation is, however, not a straightforward task; this inverse problem is the focus of our work. 
Previous work divides the diffusion curve generation problem into two steps: computation of handle geometry and assignment of handle colors.

To determine the geometry of the handles, \citet{Orzan:Diffusion:2008} and \citet{Xie:2014:Hierarchical} proposed to compute handle positions using edge detection algorithms on an input raster image.
\citet{Dai2013:Automatic} used image segmentation boundaries as handles.
\citet{Zhao:2018:InverseDiffusion} performed a PDE-constrained shape optimization using a finite element method:
 they used an isocontour of an image's per-point squared error field as the initial curve and optimized the shape of the curve by minimizing the residual error to find the optimal placement of handles. 
All of these previous approaches share the assumption that the target image is represented as a function (typically a raster image) that contains no unwanted errors.
Thus, using a Monte Carlo-rendered image as the target requires producing a fully converged image with negligible noise at a specific resolution of choice.  
The previous algorithms would otherwise fit the output vector image to reproduce any noise present in the unconverged input image. By contrast, our method \emph{directly} utilizes Monte Carlo samples from a renderer as its primary input.
Our method is similar in spirit to the work of \citet{Zhao:2018:InverseDiffusion}, in the sense that both methods solve a PDE-based optimization; however, our method employs a differential boundary element method instead, without needing complex and often fragile meshing operations during the optimization. This approach allows for robust optimization in the presence of noise from the Monte Carlo renderer.

Once the handle geometry is computed, previous work assigns handle colors either by sampling the colors of the target image locally around the handles~\cite{Orzan:Diffusion:2008, Zhao:2018:InverseDiffusion} or by solving a PDE-based least-squares fitting problem~\cite{Jeschke:2011:Estimating, Xie:2014:Hierarchical}.
In our formulation, the geometry optimization step itself already encodes color information, and we avoid a separate color optimization step.\looseness=-1

In contrast to approaches that follow a two-step geometry–color optimization framework, recent work on differential Monte Carlo PDE solvers~\cite{Miller:DiffWoS:2024, Yu:DiffWoS:2024, Yilmazer2024Solving} explores joint optimization of boundary values and geometry.
While these methods seem promising for some specific applications, they often face challenges due to slow convergence, primarily because of the high variance in estimating the normal derivatives of the solution on boundaries, which are necessary for computing the derivatives of interest.
In our preliminary experiments, we observed that this variance renders the direct application of such a method to our problem setting impractical.
Additionally, while forward Monte Carlo PDE solvers for problems involving double-sided boundaries are available~\cite{Sawhney:WoS:2020, Sawhney:WoSt:2023}, it is unclear how well the differential variants extend to inverse problems of this type.
Moreover, although our approach and Monte Carlo PDE solvers are both based on integral equation formulations, we adopt a deterministic BEM. This avoids the noise inherent in stochastic PDE solvers and allows us to optimize double-sided diffusion curve handles more directly and efficiently.

Tangentially, \citet{Li:2020:DiffVectorGraphics} proposed a differentiable rasterization framework for vector graphics. In addition to being limited by a specific resolution, their method does not support topology changes, such as the pruning of handles, and requires an additional differential Poisson equation solver to extract diffusion curves.

\section{(Differential) Boundary Element Method}
\label{sec:bem}
The core of our method is the (differential) BEM solver for diffusion curves. 
In addition to work that applies BEM to diffusion curve problems, BEM and related techniques have also been explored in graphics for a variety of applications, including linear elasticity~\cite{James1999:ArtDefo, James2003:Multires, Sugimoto2022:SurfaceOnly, Hahn2015:fracture, Hahn2016:fracture, Zhu2015:fracture}, liquid simulation~\cite{Huang2020:ferrofluids, Huang2021:Ships, Da2016:liquids, Keeler2015:Ocean, Ni2024:IoB}, sound simulation~\cite{Chadwick2009:harmonic, Zheng2010:rigid, Zheng2009:harmonic, James2006:precomputed, Umetani2016:Printone, Chew2025:Acoustic}, and various geometry processing tasks~\cite{Chen2023:Somigliana, Lipman2008:Green, Solomon2017:Octahedral, Wang2013:Harmonic, Levi2016:Convexity}.
Among these, only the works of \citet{Umetani2016:Printone} and \citet{Chew2025:Acoustic} address an inverse problem, optimizing object geometry to achieve desired acoustic properties. However, both approaches are specialized to their respective settings and do not generalize straightforwardly to our problem.

Looking beyond computer graphics, the computational mechanics community has explored BEM-based shape optimization since the 1980s, particularly for structural mechanics and heat transfer problems. \citet{Kane1988:Design-Sensitivity} provide a comprehensive overview of early work in this area. These methods typically focus on computing accurate sensitivity information to guide shape updates, often assuming clean boundary representations or employing implicit representations, such as level set methods~\cite{Abe2007}, to obtain clean boundaries and manage geometric complexity. While our work shares the goal of computing accurate differentials, we additionally seek to robustly optimize explicitly represented diffusion curve handles. To this end, we utilize a smoothed version of the fundamental solution, enabling stable and direct gradient computation even in the presence of unstructured, sometimes intersecting, double-sided boundaries. Moreover, our choice of optimization variable, which avoids repeated factorization of the linear system during optimization, has not been presented in previous work, to our knowledge.

\subsection{Forward Solver}\label{sec:bem_forward}
BEM solvers for diffusion curves are well-established in graphics. In this section, we summarize the relevant background and outline the specific formulation choices made for our method.
\subsubsection{Boundary integral equation}
The boundary element method reformulates a given PDE as a \emph{boundary integral equation} (BIE), which is then solved by discretizing only the boundary and not the domain interior. For the Laplace equation with an outer (Neumann) boundary $\partial\domain$ and double-sided boundaries $\handleset(\parm)$, the corresponding boundary integral equation for $\sol(\vecx)$ is~\cite{Bang:Diffusion:2023, Chen:2025:Lightning, Sun:2012:Texture, Sun:2014:FastMultipole, Gronde:DiffusionCurves:2010}:
\begin{equation}\label{eq:bie}
\alpha(\vecx)\sol(\vecx) \\
= -\int_{\partial\domain} \frac{\partial G}{\partial \vecn_\vecy}(\vecx, \vecy)\sol(\vecy) \dy
 + \handleterm(\vecx; \parm)
\end{equation}
where $\alpha(\vecx) = 1$ for $\vecx\in\domain$ and $\alpha(\vecx) = 0.5$ for $\vecx$ on a smooth domain boundary and $\vecn_\vecy$ is the outward normal direction. We have omitted the additional term $\int_{\partial\domain} G(\vecx, \vecy)\frac{\partial \sol}{\partial \vecn_\vecy}(\vecy) \dy$ that typically appears in the BIE because the boundary condition in \cref{eq:neumann} ensures that this term is zero.
Here, $G(\vecx, \vecy)$ is the fundamental solution for the Laplace equation. In 2D,
\begin{equation}\label{eq:fundsol}
    \begin{split}
        G(\vecx, \vecy) &= -\frac{1}{2\pi} \log r\quad\text{and}\quad \frac{\partial G}{\partial \vecn_\vecy}(\vecx, \vecy) = -\frac{1}{2\pi} \frac{\vecn_\vecy\cdot(\vecy - \vecx )}{r^2},
    \end{split}
\end{equation}
where $r=\lVert\vecy-\vecx\rVert_2$.
The term $\handleterm(\vecx;\parm)$ accounts for the contributions from double-sided ("jump") boundaries:
\begin{equation}\label{eq:handleterm}
\handleterm(\vecx;\parm)= -\int_{\handleset(\parm)} \frac{\partial G}{\partial \vecn_\vecz}(\vecx, \vecz)\weightd(\vecz)  \dz + \int_{\handleset(\parm)}   G(\vecx, \vecz)\weightc(\vecz) \dz
\end{equation}
where $\vecn_\vecz$ is the normal of the handle, and $\weightd$ and $\weightc$ represent the jumps in the color and the normal derivative of the color, respectively: $\weightd(\vecz;\parm)= \handlecolor^+(\vecz;\parm) - \handlecolor^-(\vecz;\parm)$ and $\weightc(\vecz) =\frac{\partial\handlecolor}{\partial\vecn_\vecz}^+(\vecz;\parm) - \frac{\partial\handlecolor}{\partial\vecn_\vecz}^-(\vecz;\parm)$. Notice that this formulation specifies these jump values directly, in contrast to the more common double-sided Dirichlet color specification of \cref{eq:dirichlet}.
The term involving $\weightd$ introduces discontinuities in the solution across the handle, while the term involving $\weightc$ contributes smoothly across it, only introducing discontinuity in the gradient of the solution~\cite{Gronde:DiffusionCurves:2010}. %

\subsubsection{Compatibility condition}\label{sec:compatibility}
To ensure that the boundary integral equation~(\cref{eq:bie}) is valid, the boundary data must satisfy a compatibility condition: the net flux of the normal derivative across the boundary must vanish. Since we assume zero Neumann conditions on $\partial\domain$, we need only consider the flux from the handles:
$\int_{\handleset(\parm)} \weightc(\vecz) \dz = 0.$

\subsubsection{Solution of BIE via discretizations}\label{sec:bem_discretization}
We now describe how to discretize~\cref{eq:bie} and compute the solution $\sol$ within the domain $\domain$. As before, we assume zero Neumann boundary conditions on $\partial\domain$, and jump conditions on the handles $\handleset(\parm)$ defined via weights $\weightd$ and $\weightc$ that satisfy the compatibility condition.
We first consider applying \cref{eq:bie} at domain boundary points $\vecx\in\partial\domain$, to solve for the unknowns $\sol$ on the boundary.
We discretize $\partial\domain$ into small segments (e.g., shorter than $1/256$ of the image width in our implementation), assuming $\sol$ is piecewise constant over each. Applying the integral equation at each segment yields a linear system. Using a Galerkin discretization, the equation for segment $i$ becomes
\begin{equation}\label{eq:bie_galerkin}
\left(\int_{\Gamma_i}\frac{1}{2}\dx\right) \discrete{u}_i + \sum_j  \left(\int_{\Gamma_i} \int_{\Gamma_j}\frac{\partial G}{\partial \vecn_\vecy}(\vecx, \vecy) \dy\dx \right) \discrete{u}_j
 = \int_{\Gamma_i}\handleterm(\vecx;\parm) \dx,
\end{equation}
where $\discrete{u}_i$ is the constant solution on the $i$-th segment $\Gamma_i$, and the sum over $j$ includes all segments.
We evaluate all integrals in~\cref{eq:bie_galerkin} using 3-point Gaussian quadrature.
The integral kernel $\frac{\partial G}{\partial\vecn_\vecy}$ has a singularity at $\vecx = \vecy$, i.e., when $\Gamma_i = \Gamma_j$.
We assume that each segment $\Gamma_i$ is a line segment, so for this coincident element case, $\vecn_\vecy\cdot(\vecy - \vecx ) = 0$, and the integral evaluates to zero analytically.
While our implementation aims for simplicity, it should be possible to adopt more accurate yet complex discretizations and analytical evaluation or quadrature rules (e.g., as explained by \citet{Bang:Diffusion:2023}) if desired.\looseness=-1

To compute $\handleterm(\vecx;\parm)$, we similarly discretize each handle into small segments and apply Gaussian quadrature. As the integrands contained in $\handleterm(\vecx;\parm)$ have singularities at $\vecx = \vecz$, for computational purposes, we regularize the fundamental solution (\cref{eq:fundsol}) contained in this term by replacing $r$ with $\sqrt{r^2 + \varepsilon^2}$, where $\varepsilon$ is a small constant, similar to \citet{Chen:2024:Lightning}.

By assembling \cref{eq:bie_galerkin} for each segment, we obtain a global linear system $\mathbf{A}\discrete{u} = \discrete{f}$. Solving this system yields the boundary value $\discrete{u}_i$ for each segment.
However, the system exhibits a one-dimensional null space, because the domain boundaries satisfy Neumann conditions and the handles only have jumps ($\weightd$ and $\weightc$) prescribed; thus the solution is defined only up to an additive constant.
This null space corresponds to constant functions over the domain. To obtain a unique solution, we solve the system using singular value decomposition (SVD), discarding the smallest singular value to remove this ambiguity. This effectively projects the solution onto the subspace orthogonal to constants, enforcing a zero-mean constraint over the domain. The compatibility condition ensures that the right-hand side lies in the range of the operator, guaranteeing solvability.

Once we compute the boundary values $\discrete{u}_i$, we recover the solution $\sol(\vecx)$ at any point $\vecx \in \domain$ by evaluating \cref{eq:bie} using the discretized boundary data. Because the BEM solve only provides a solution up to a constant, we separately add back a prescribed constant per domain to obtain the final color at each evaluation point. As the BEM solve provides the zero-mean solution, this given constant should represent the mean color of the domain.

The most computationally expensive step in this process is the factorization of the dense system matrix $\mathbf{A}$, which scales cubically with the number of outer boundary segments. However, when solving problems with a fixed domain boundary $\partial \Omega$ for multiple sets of handles, we can factorize $\mathbf{A}$ once at the beginning and then reuse this factorization for all subsequent solves. This may seem counterintuitive and different from previous applications of BEM for diffusion curves, but it arises from our particular formulation: we assume that the handles' jump values are directly specified, rather than, e.g., the colors on either side of the handles, as in traditional formulations. Consequently, the matrix $\mathbf{A}$ depends only on the pairwise geometric relationship of the outer, Neumann boundary segments, and all handle-related parameters are incorporated explicitly in the right-hand side vector $\mathsf{f}$, decoupling them from the degrees of freedom in the linear system. This separation allows the factorization of $\mathbf{A}$ to be reused across multiple solves, rendering its cost negligible in practice.

\subsubsection{Recovering double-sided colors} Thus far, we have described how to evaluate the solution of the BIE at any point $\vecx \in \domain$, given the color jump $\weightd$ and the normal derivative jump $\weightc$ on each handle. Although we employ the jump formulation rather than the explicit double-sided color specification from \cref{eq:dirichlet}, we can easily recover the latter. Specifically, at points $\vecx \in \handleset(\parm)$, we evaluate the reconstructed color $\sol(\vecx)$ using the same procedure as for interior points. This yields the average color $\handlecolor_\text{avg}$ across the handle, as noted by \citet{Bang:Diffusion:2023}. Together with the prescribed color difference $\weightd$, we can reconstruct the color on each side of the handle as $\handlecolor_\text{avg} \pm \weightd/2$.

\subsection{Differential Solver}
Our goal is to estimate the handle parameters $\parm$ using gradient-based optimization so that the reconstructed image $\sol(\vecx)$ matches a target image. This requires computing the Jacobian of the solution with respect to $\parm$, denoted $\frac{\partial \sol}{\partial \parm}$. While researchers have long used boundary element methods (BEM) to solve forward problems, to our knowledge, no prior work in graphics has applied BEM to directly compute solution differentials for optimization of diffusion curves.

\subsubsection{Smoothing discontinuity} In our setup, the outer boundary $\partial\domain$ remains fixed and satisfies zero Neumann conditions, while the interior handle boundaries $\handleset(\parm)$ vary with $\parm$. These handles introduce color discontinuities in the reconstructed image, which pose challenges for gradient-based methods. Standard optimization techniques typically assume a smooth objective function, and discontinuities often require special handling~\cite{Zhao2020:PBDR}.
The source of discontinuity in our formulation is the first term in \cref{eq:handleterm}, which involves the strongly singular kernel $\frac{\partial G}{\partial \vecn_\vecz}$. We regularize this kernel, as in the forward solve (\cref{sec:bem_discretization}). While introducing some approximations to the original formulation, this approach not only improves the robustness of numerical integral evaluation but also eliminates discontinuities in the reconstructed color field. 
Increasing the regularization parameter $\varepsilon$ for this term further smooths out the discontinuity, which improves optimization stability.

\subsubsection{Differential BIE}
With this smoothing in place, we differentiate the BIE in \cref{eq:bie} with respect to $\parm$, yielding what we term a \emph{differential BIE}:
\begin{equation}\label{eq:diffbie}
\alpha(\vecx)\frac{\partial \sol}{\partial \parm}(\vecx)
= -\int_{\partial\domain} \frac{\partial G}{\partial \vecn_\vecy}(\vecx, \vecy)\frac{\partial \sol}{\partial \parm}(\vecy) \dy +
\frac{\partial \handleterm}{\partial \parm}(\vecx; \parm_{\text{curr}}).
\end{equation}
Since the outer boundary $\partial\domain$ does not depend on $\parm$, the differential BIE preserves the same structure as the original BIE. In \cref{eq:diffbie}, the unknown solution $\sol$ in \cref{eq:bie} is replaced by its derivative $\frac{\partial \sol}{\partial \parm}$, and likewise $\handleterm$ is replaced by its derivative. 
As we assume $\handleterm$ has been smoothed by regularization, the numerical evaluation of $\frac{\partial \handleterm}{\partial \parm}$ does not suffer from higher-order singularities that would be introduced otherwise.
To evaluate this derivative, we express each integral in \cref{eq:handleterm} as a sum over individual handles. Each handle is represented as a line segment, and we reformulate its integral in a parametric form, where the handle geometry is defined by two endpoints interpolated via a scalar parameter.
This change of variables makes the integration domain independent of $\parm$, allowing us to differentiate it with respect to all parameters and evaluate it using a simple quadrature rule.\looseness=-1

\subsubsection{Discretization}
To evaluate $\frac{\partial \sol}{\partial \parm}$ at any point $\vecx \in \domain$, we follow the same procedure as in the forward solve. Given the current parameters $\parm_{\text{curr}}$, we can evaluate (the regularized version of) $\frac{\partial \handleterm}{\partial \parm}(\vecx; \parm_{\text{curr}})$ at any $\vecx\in\partial\domain$.
Using a constant element Galerkin method, we obtain a discretized system similar to \cref{eq:bie_galerkin}, which we solve to compute the discretized boundary values of $\frac{\partial \sol}{\partial \parm}$. Once we obtain these values, we can evaluate $\frac{\partial \sol}{\partial \parm}$ at any point in the domain via \cref{eq:diffbie}.

Critically, the matrix system for this Jacobian solve is identical to the one used for the forward solve, and it remains unchanged throughout the optimization.
During the Jacobian solve, the right-hand side becomes a collection of vectors, each of which corresponds to the derivative of $\handleterm$ with respect to each parameter, instead of a single vector for $\handleterm$ itself.
Similar to the forward solver, all optimizable parameters and their derivatives appear in the right-hand (stacked) vector, $\frac{\partial \mathbf{f}}{\partial \theta}$.
Thus, we can factorize the matrix only once at the beginning and reuse it; we can efficiently compute both the solution and its differentials by solving the system with different right-hand sides.\looseness=-1

\section{Diffusion Curve Optimization}
\label{sec:optimization}
Equipped with the forward and differential BEM solvers, we now develop our method to construct a diffusion curve representation of an image directly from the noisy samples of a Monte Carlo renderer.
We formulate our task as an optimization problem to find the parameters of the handles.
After the initialization of handles (\cref{sec:initialization}), we optimize the handles with the primary objective of minimizing the least squares reconstruction loss against the target image $\itar(\vecx)$ over the whole image $\imagedomain$,
\begin{equation}\label{eq:squaredl2loss}
    \loss(\parm) = \frac{1}{2}\int_\imagedomain \lVert \ireconst(\vecx; \parm) - \itar(\vecx) \rVert^2_2\,d\vecx,
\end{equation}
with additional regularization losses (\cref{sec:regularization}). 
We assume that the target image $\itar(\vecx)$ is available only through a Monte Carlo renderer that produces random samples at a given arbitrary spatial location. Specifically, for any query point $\vecx \in \imagedomain$, the renderer provides noisy samples whose \emph{expectation} equals the true target pixel intensity at the sampled image location.
We assume that we operate on the \emph{linear} color space for $\ireconst$ and $ \itar$. 
To perform a gradient-based optimization with the least squares loss, we use the differential BEM (\cref{sec:bem}) to estimate $\ireconst(\vecx; \parm)$ and $\frac{\partial \ireconst}{\partial \parm}(\vecx; \parm)$. The optimization follows the Levenberg-Marquardt framework (\cref{sec:levenberg}), applied within image subdomains (\cref{sec:domaindecomp}).

\subsection{Domain decomposition}\label{sec:domaindecomp}
\newlength{\origcolumnsep}
\setlength{\origcolumnsep}{\columnsep}
The input to our method is a 3D scene, and we use object contour information to define image-space subdomains $\domain_i \subset \mathbb{R}^2$. We then solve the optimization problem within each subdomain, rather than solving it globally over the entire image domain $\imagedomain$.
To determine such subdomains, we project objects or primitives in the 3D scene, as specified by the user, onto the 2D image plane, allowing overlaps. We extract exterior contours using a simple ray intersection-based check, followed by simple post-processing to refine the boundaries. 
\setlength{\columnsep}{5pt}
\begin{wrapfigure}[7]{r}{0.3\linewidth}
\vspace{-\intextsep}
\includegraphics[width=\linewidth]{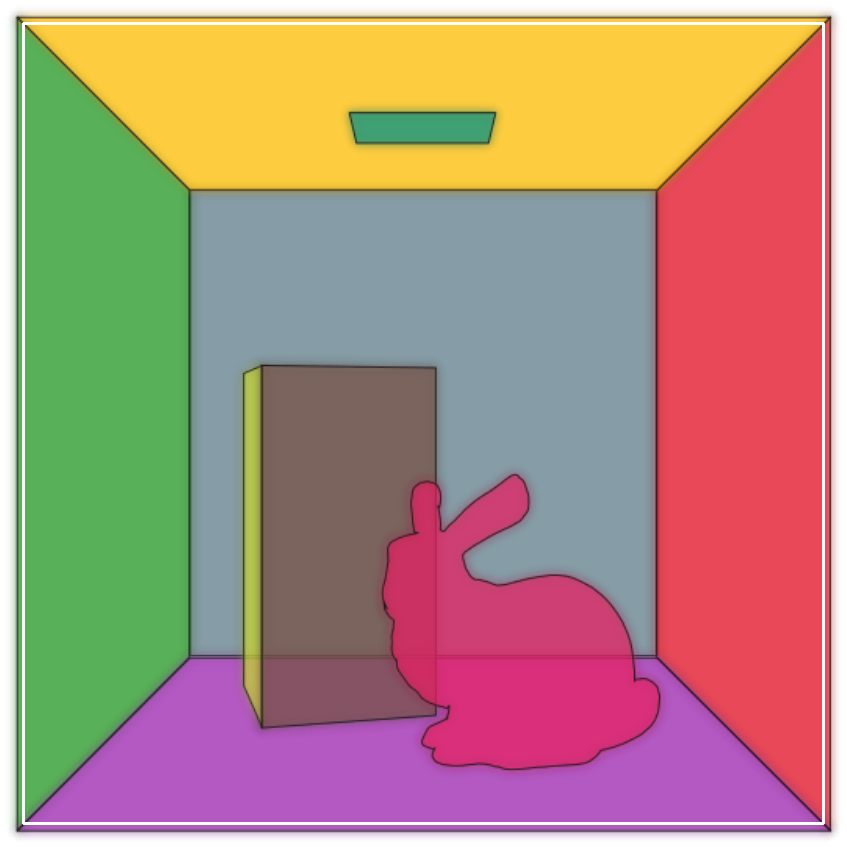}
\end{wrapfigure}
Note that a subdomain $\domain_i$ may include parts of the projected geometry that are occluded by other objects or lie outside $\imagedomain$, as long as some portion of the subdomain is visible from the camera (see inset).
We make this choice of allowing overlaps to simplify domain boundary extraction and allow the handles to influence the colors of the subdomain, even from or through the occluded part of the subdomain.  
This domain decomposition approach is effective because the subdomain boundaries often correspond to sharp color discontinuities in the image, and we can handle these discontinuities by placing zero Neumann boundaries without relying on optimization.
Given an image space position $\vecx\in\imagedomain$, we also assume that we can associate it with a unique subdomain by casting a ray from the camera and checking which object or primitive it hits first in the input 3D scene. 
Within each subdomain, our task is to find the parameters for diffusion curve handles that minimize the reconstruction loss.
Apart from associating each image-space Monte Carlo sample to a corresponding subdomain, each subdomain is treated separately in our computation, and each handle is assigned to one initial subdomain and remains associated with the same subdomain throughout optimization.
Within each subdomain, each handle is represented using the positions of the two endpoints in $\domain_i$, together with the color and derivative jumps, which are assumed to be constant over each handle.
Additionally, we include one more optimizable color parameter per subdomain, which represents the spatially constant mean color of that subdomain.

\setlength{\columnsep}{\origcolumnsep}

\subsection{Initialization} \label{sec:initialization}
At the beginning of the optimization, we place 500 handles in the image space. The initial positions of the handles are determined by sampling from a quasirandom number sequence~\cite{Joe:2008:ConstructingSobol}. Specifically, we first sample an endpoint, and then we sample the second endpoint in such a way that the initial length of each handle is set to $1/20$ of the image width. We then snap the second endpoint to the same subdomain as the first, so each handle strictly belongs to one subdomain without ambiguity. Note that this handle initialization procedure frequently yields configurations with many handles intersecting one another. Nevertheless, our method can robustly optimize the parameters without problems.

We also initialize the SVD factorization required for BEM solves. As described in \cref{sec:bem}, we discretize the fixed outer boundary $\partial\domain_i$ into small segments, assemble the dense system matrix, and precompute its SVD. Then, we can reuse the same factorization throughout the optimization, avoiding repeated expensive matrix factorizations.

\subsection{Levenberg--Marquardt Optimization}
\label{sec:levenberg}

Our optimization problem involves a relatively small number of parameters and aims to minimize the least squares loss (\cref{eq:squaredl2loss}). This setting makes the Levenberg–Marquardt method~\cite{Nocedal:2006:Optimization} a natural choice. 
Unlike more common stochastic gradient descent approaches, Levenberg–Marquardt uses both the Jacobian and an approximate Hessian of the loss, and we can expect faster and more stable convergence with minimal hyperparameter tuning, especially when the parameter scales are nonuniform, as in our problem with both geometry and jump weight parameters.

We can derive the update rule by linearizing $\ireconst(\vecx; \parm)$ around the current parameters $\parm_i$ and setting the gradient of the loss to zero. This yields the following parameter update:
\begin{equation}\label{eq:optimstep}
\begin{split}
    &\parm_{i+1}\\
    &=\parm_i -{\biggl\{\damping\mathbf{D} + \underbrace{\int_\imagedomain  J_i\transpose (\vecx)  J_i(\vecx)\,d\vecx \biggl\}}_{\text{approximate loss Hessian}}}^{-1}
    \underbrace{\int_\imagedomain  J_i\transpose (\vecx) \left\{\ireconst(\vecx; \parm_i) - \itar(\vecx)\right\} d\vecx}_{\text{loss Jacobian}},
\end{split}
\end{equation}
where $J_i(\vecx) = \frac{\partial \ireconst}{\partial \parm}(\vecx; \parm_i)$ is the image-space Jacobian, $\mathbf{D}$ is the diagonal matrix that we get from the diagonals of the first integral, and $\damping$ is a damping parameter.
This update approximates a Newton step using only first-order information. The first integral approximates the Hessian of the loss, and the second corresponds to the gradient. We estimate both using Monte Carlo integration, as we describe later.
The damping term $\damping\mathbf{D}$ essentially automatically adjusts the learning rate of the optimization, and we dynamically update the damping parameter $\damping$ by checking the change of loss at each step of the optimization, following \citet{transtrum2012improvementslevenbergmarquardtalgorithmnonlinear}.
Because the approximate Hessian is positive definite, we solve the resulting linear system at each iteration using Cholesky factorization. Furthermore, since handles in a subdomain $\domain_i$ only influence the reconstructed color within that subdomain, the approximate Hessian has a block-diagonal structure when the handles are ordered by subdomains. Therefore, we store the approximate Hessian and Jacobian independently for each subdomain and solve the systems separately. We repeat this parameter update process based on \cref{eq:optimstep} iteratively until convergence.

\paragraph*{Monte Carlo integration}
To estimate the integrals in \cref{eq:optimstep}, in each step of optimization, we draw $M_\text{H}$ samples from a probability density function $p_\text{H}(\vecx)$ to approximate the loss Hessian, and $M_\text{J}$ samples from a (potentially different) density $p_\text{J}(\vecx)$ to approximate the loss Jacobian:
\begin{equation}\label{eq:losshessian}
\int_\imagedomain  J_i\transpose (\vecx)  J_i(\vecx)\,d\vecx 
\approx \frac{1}{M_\text{H}}\sum_{j=1}^{M_\text{H}} \frac{J_i\transpose (\vecx_j)  J_i(\vecx_j)}{p_\text{H}(\vecx_j)}, 
\end{equation}
\begin{equation}\label{eq:lossjacobian}
\int_\imagedomain  J_i\transpose (\vecx) \left\{\ireconst(\vecx; \parm_i) - \itar(\vecx)\right\} d\vecx 
\approx \frac{1}{M_\text{J}}\sum_{j=1}^{M_\text{J}} \frac{J_i\transpose (\vecx_j) \left\{\ireconst(\vecx_j; \parm_i) - \itar(\vecx_j)\right\}}{p_\text{J}(\vecx_j)}.
\end{equation}
We sample 100,000 new points uniformly randomly in each step of optimization and use them to evaluate both integrals in our implementation.
We evaluate both $\ireconst(\vecx; \parm_i)$ and its Jacobian $J_i(\vecx)$ analytically using the differential BEM, and replace $\itar(\vecx_j)$ by its stochastic estimation with Monte Carlo rendering (with $128$ samples per $\vecx_j$ in our implementation).
We observed that excessive intensity values in our optimization can lead to instability likely due to our choice of the (absolute) least-squares loss. Therefore, we clamp the estimate of $\itar(\vecx_j)$ to a maximum value of $1$.
In the damping parameter update, we use the samples used for optimization to also check the change in loss.

\paragraph*{Compatibility condition enforcement and mean color.}
As discussed in \cref{sec:compatibility}, the handle parameters must satisfy the compatibility condition. A naive parameter update using the rule in \cref{eq:optimstep} can violate this constraint. To enforce compatibility during optimization, we project both the approximate Hessian and the loss Jacobian onto the subspace of feasible solutions under the assumption of fixed endpoints. In our parameterization, this constraint corresponds to requiring that the weighted sum of the handle values $\weightc$ equals zero, where the weights are given by the lengths of the associated handles. 
Since this projection is based on the fixed-endpoint assumption, it may not fully eliminate errors introduced by the parameter update. To correct for any remaining violations, we reproject the $\weightc$ values onto the constraint manifold after each update, using the updated handle endpoint positions. This two-stage projection ensures that the compatibility condition remains satisfied throughout the optimization.
For the mean color constant estimation, we take the mean of all the renderer samples generated within each domain independently from the Levenberg-Marquardt framework, to ensure a more accurate estimate without oscillation.

\subsection{Regularization}\label{sec:regularization}

In addition to the reconstruction loss (\cref{eq:squaredl2loss}), we can incorporate regularization losses to achieve qualitatively desirable diffusion curve handles. For each regularization loss term, we add its analytical Hessian and gradient to the first and second integrals in \cref{eq:optimstep}, respectively.

\paragraph*{Handle length}

Without regularization, handles tend to lengthen during optimization, but a single long handle influencing a large region of the image may not be desirable. We therefore add a regularization term that penalizes handles whose length exceeds a specified threshold. For each handle, we measure the Euclidean distance between its endpoints. If this length exceeds the threshold, we apply a quadratic penalty proportional to the amount by which it exceeds the limit. No regularization is applied if the handle length is below the threshold.

\paragraph*{Endpoint snapping}
We also encourage different handles' endpoints to be close to each other, so as to better approximate continuous boundaries. 
When two endpoints of two different line segments are closer than a certain (Euclidean distance) threshold, we decide whether to encourage them to coalesce based on two factors: the Euclidean distance between the endpoints and the directional similarity of the segments, where the latter is computed using the dot product of the line directions. 
We define a \emph{snapping score} as the weighted average of these two terms. When such a pair of endpoints gives a mutually smallest snapping score, we add a regularization loss defined as the squared Euclidean distance between the endpoints, effectively encouraging them to snap. %

\paragraph*{Sparsity}
Optionally, we can aim to represent the target image using as few handles as possible, balancing reconstruction quality and sparsity.
To achieve this, we add a small saturated sparsity regularization loss for each handle weight:
$\lambda_w t\cdot \tanh(\sqrt{\weight\cdot\weight + \epsilon^2}/t)$, where 
$\lambda_w$, $\epsilon$, and $t$ are small positive constants, and $\weight = \weightc\text{ or }\weightd$. We use a larger regularization parameter $\lambda_w$ for $\weightd$ than for $\weightc$.
Then, during the iterative optimization process, we eliminate a handle if the norms of both of its associated weights fall below a predefined threshold.
In our implementation, we make sparsity regularization optional and do not enable it by default, as we often observed that the sparsity loss compromises the reconstruction quality.\looseness=-1

\section{Results}

\begin{figure}[t]
\centering
  \begin{tikzpicture}
    \node[inner sep=0pt] (img1) at (0,0)
      {\includegraphics[width=0.25\linewidth]{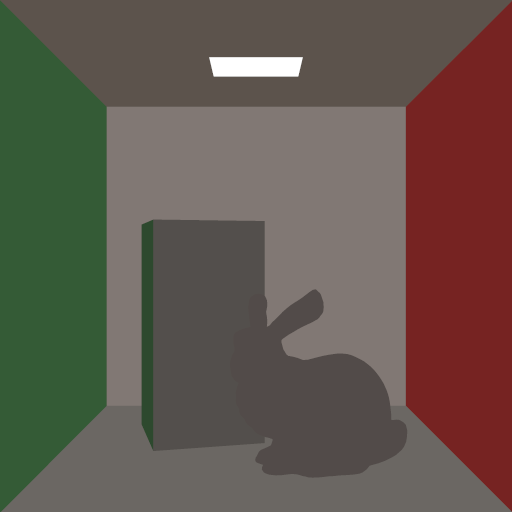}};
    \node[anchor=north west] at (img1.north west) {\footnotesize\textsf{\textcolor{white}{\shortstack[l]{Step 1\\0.036 min}}}};
  \end{tikzpicture}%
  \begin{tikzpicture}
    \node[inner sep=0pt] (img2) at (0,0)
      {\includegraphics[width=0.25\linewidth]{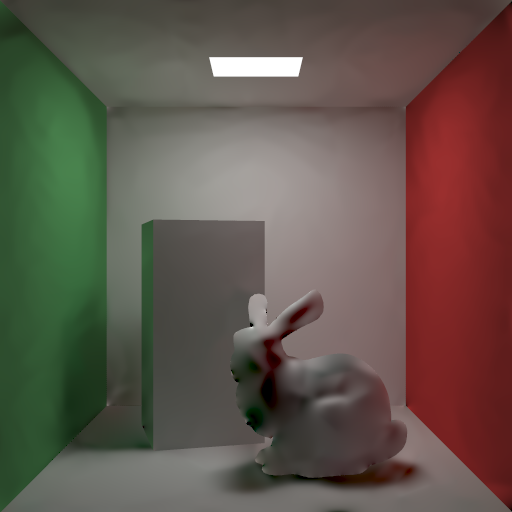}};
    \node[anchor=north west] at (img2.north west) {\footnotesize\textsf{\textcolor{white}{\shortstack[l]{25\\9.00 min}}}};
  \end{tikzpicture}%
  \begin{tikzpicture}
    \node[inner sep=0pt] (img3) at (0,0)
      {\includegraphics[width=0.25\linewidth]{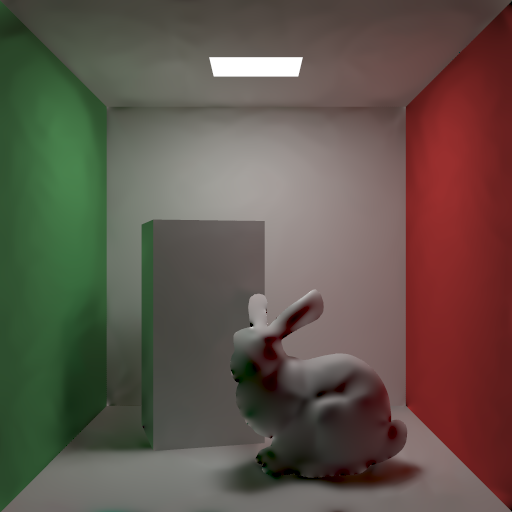}};
    \node[anchor=north west] at (img3.north west) {\footnotesize\textsf{\textcolor{white}{\shortstack[l]{50\\18.4 min}}}};
  \end{tikzpicture}%
  \begin{tikzpicture}
    \node[inner sep=0pt] (img4) at (0,0)
      {\includegraphics[width=0.25\linewidth]{Figures/results/cb2/cb2_baseline_100.png}};
    \node[anchor=north west] at (img4.north west) {\footnotesize\textsf{\textcolor{white}{\shortstack[l]{100\\37.4 min}}}};
  \end{tikzpicture}\\
\caption{Progressive optimization. 
Our method quickly captures the overall pattern of the image in a few minutes, and progressively refines the handles.}
\label{fig:progressive}
\end{figure}

\begin{figure}[t]
\centering
\includegraphics[width=0.245\linewidth]{Figures/results/cb2/cb2_baseline_100.png}%
\includegraphics[width=0.245\linewidth]{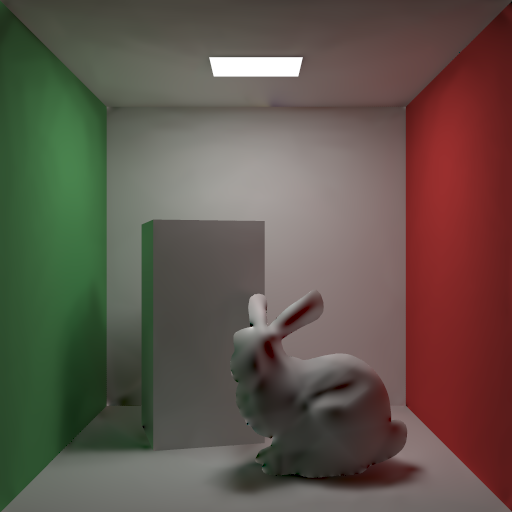}%
\includegraphics[width=0.245\linewidth]{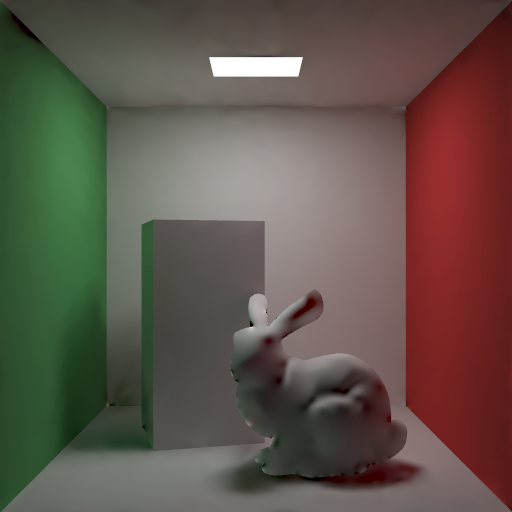}%
\includegraphics[width=0.245\linewidth]{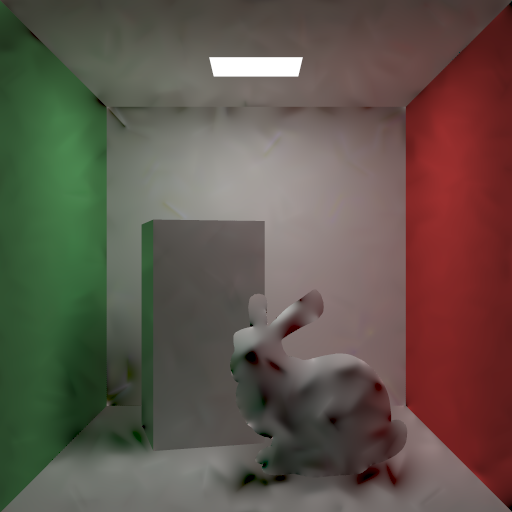}\\
\includesvg[width=0.245\linewidth]{Figures/results/cb2/cb2_baseline_100.svg}%
\includesvg[width=0.245\linewidth]{Figures/results/cb2/cb2_nolengthreg_100.svg}%
\includesvg[width=0.245\linewidth]{Figures/results/cb2/cb2_nosnapping_100.svg}%
\includesvg[width=0.245\linewidth]{Figures/results/cb2/cb2_wopos_100.svg}\\%
\begin{minipage}[b]{0.245\linewidth}\centering\textsf{\small Baseline}\end{minipage}%
\begin{minipage}[b]{0.245\linewidth}\centering\textsf{\small No Length Reg.}\end{minipage}%
\begin{minipage}[b]{0.245\linewidth}\centering\textsf{\small No Snapping}\end{minipage}%
\begin{minipage}[b]{0.245\linewidth}\centering\textsf{\small No Pos. Update}\end{minipage}\\%
\begin{minipage}[b]{0.245\linewidth}{\small \textsf{RMSE: } $0.0138$}\end{minipage}%
\begin{minipage}[b]{0.245\linewidth}\centering{\small $0.0134$}\end{minipage}%
\begin{minipage}[b]{0.245\linewidth}\centering{\small $0.0138$}\end{minipage}%
\begin{minipage}[b]{0.245\linewidth}\centering{\small $0.0198$}\end{minipage}%

\caption{Handle positions. Compared to the baseline, disabling either the handle length regularization or the endpoint snapping regularization results in handles that become arbitrarily long or are distributed without forming coherent structures, despite achieving a slight improvement in reconstruction error. When the handle position update is disabled, the output retains artificial patterns introduced by the initial handle configuration, and the reconstruction error is higher.}
\label{fig:positions}
\end{figure}

\begin{figure*}
\centering
\begin{minipage}[b]{0.12\linewidth}\centering\textsf{\small Target\\\;}\end{minipage}
\includegraphics[width=0.2\linewidth]{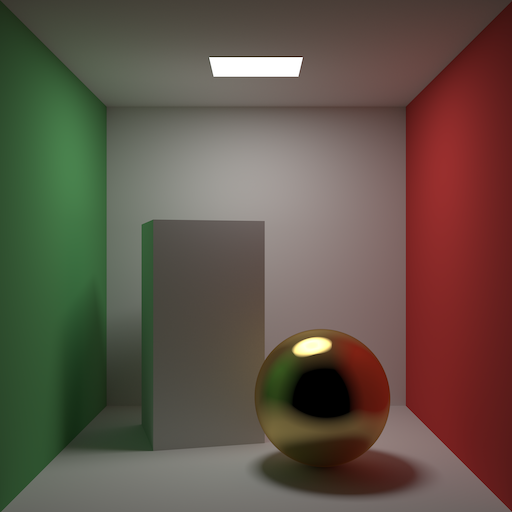}
\includegraphics[width=0.2\linewidth]{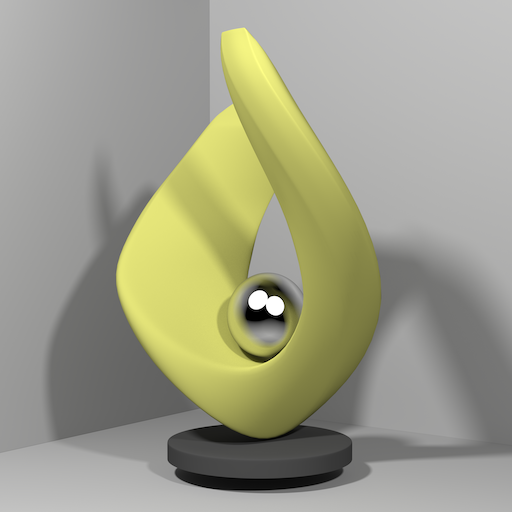}
\includegraphics[width=0.2\linewidth]{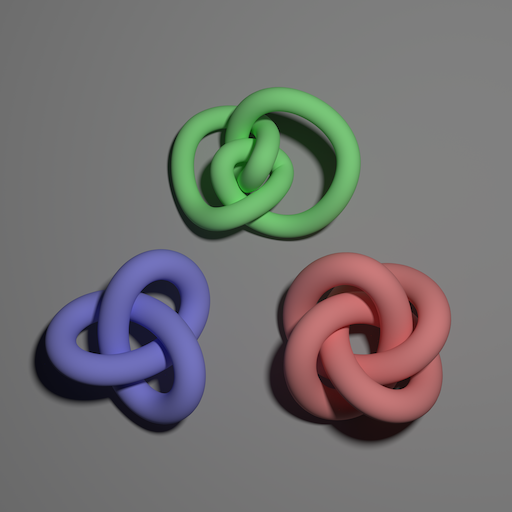}
\includegraphics[width=0.2\linewidth]{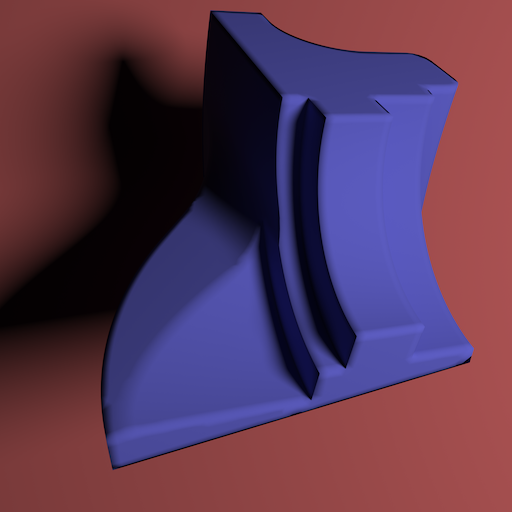}

\begin{minipage}[b]{0.12\linewidth}\centering\textsf{\small Reconstructed\\\;}\end{minipage}
\includegraphics[width=0.2\linewidth]{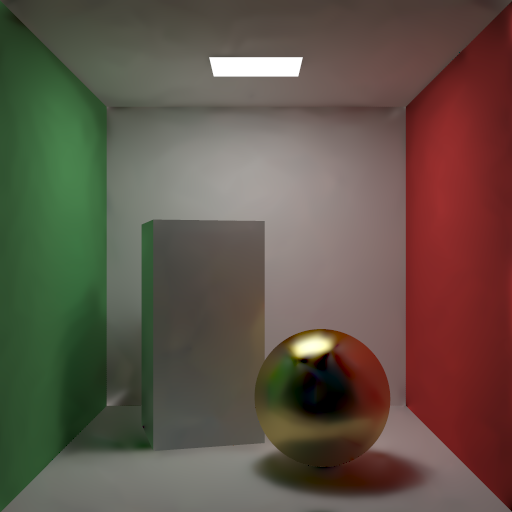}
\includegraphics[width=0.2\linewidth]{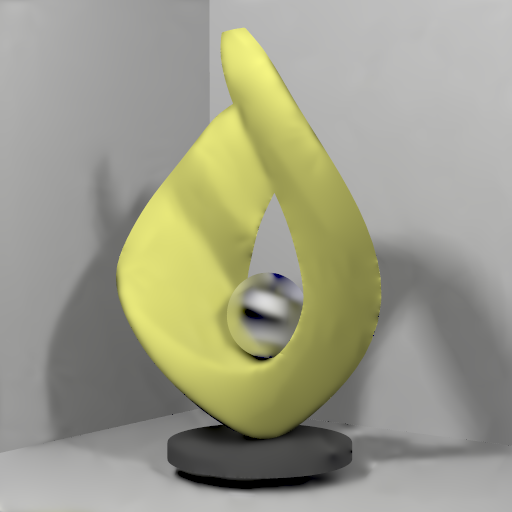}
\includegraphics[width=0.2\linewidth]{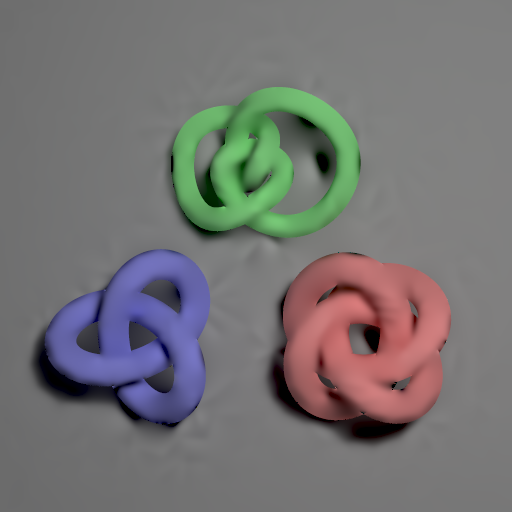}
\includegraphics[width=0.2\linewidth]{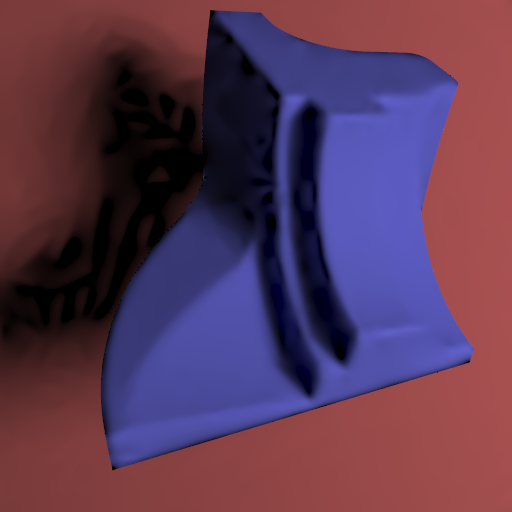}

\begin{minipage}[b]{0.12\linewidth}\centering\textsf{\small Diffusion\\Curves\\\;}\end{minipage}
\includesvg[width=0.2\linewidth]{Figures/results/cb4_100.svg}
\includesvg[width=0.2\linewidth]{Figures/results/sculpture_100.svg}
\includesvg[width=0.2\linewidth]{Figures/results/knots_100.svg}
\includesvg[width=0.2\linewidth]{Figures/results/fandisk_100.svg}

\begin{minipage}[b]{0.12\linewidth}\centering\textsf{\small Error Curve}\end{minipage}
\begin{minipage}[b]{0.2\linewidth}
\centering
    \resizebox{1.0\linewidth}{!}{
    \begin{tikzpicture}
    \begin{axis}[width=\linewidth,height=0.8\linewidth,
    ymode=log, log basis y=10,
    grid=both, scale only axis,
    xtick={0, 25, 50, 75, 100},
    ytick={0.1, 0.05623413251, 0.031622776601, 0.0177827941, 0.01},
    xmin=0,
    xmax=100,
    axis background/.style={fill=gray!5},
    every axis x label/.style={at={(0.5, -0.1)},anchor=north, font=\sffamily\bfseries\small},
    every axis y label/.style={at={(0, 1)},anchor=north east, font=\sffamily\bfseries\small},
    ylabel=RMSE,
    xlabel=Step
    ]
    \addplot+[line width=2pt,mark=none,color=mycolor1]table[]{Figures/results/cb4_step_vs_rmse_tonemapped_abs.tex};
    \end{axis}
    \end{tikzpicture}}
\end{minipage}
\begin{minipage}[b]{0.2\linewidth}
\centering
    \resizebox{1.0\linewidth}{!}{
    \begin{tikzpicture}
    \begin{axis}[width=\linewidth,height=0.8\linewidth,
    ymode=log, log basis y=10,
    grid=both, scale only axis,
    xtick={0, 25, 50, 75, 100},
    ytick={0.177827941,0.1, 0.05623413251, 0.031622776601, 0.0177827941, 0.01},
    xmin=0,
    xmax=100,
    axis background/.style={fill=gray!5},
    every axis x label/.style={at={(0.5, -0.1)},anchor=north, font=\sffamily\bfseries\small},
    every axis y label/.style={at={(0, 1)},anchor=north east, font=\sffamily\bfseries\small},
    ylabel=RMSE,
    xlabel=Step
    ]
    \addplot+[line width=2pt,mark=none,color=mycolor1]table[]{Figures/results/sculpture_step_vs_rmse_tonemapped_abs.tex};
    \end{axis}
    \end{tikzpicture}}
\end{minipage}
\begin{minipage}[b]{0.2\linewidth}
\centering
    \resizebox{1.0\linewidth}{!}{
    \begin{tikzpicture}
    \begin{axis}[width=\linewidth,height=0.8\linewidth,
    ymode=log, log basis y=10,
    grid=both, scale only axis,
    xtick={0, 25, 50, 75, 100},
    ytick={0.1, 0.05623413251, 0.031622776601, 0.0177827941, 0.01},
    xmin=0,
    xmax=100,
    axis background/.style={fill=gray!5},
    every axis x label/.style={at={(0.5, -0.1)},anchor=north, font=\sffamily\bfseries\small},
    every axis y label/.style={at={(0, 1)},anchor=north east, font=\sffamily\bfseries\small},
    ylabel=RMSE,
    xlabel=Step
    ]
    \addplot+[line width=2pt,mark=none,color=mycolor1]table[]{Figures/results/knots_step_vs_rmse_tonemapped_abs.tex};
    \end{axis}
    \end{tikzpicture}}
\end{minipage}
\begin{minipage}[b]{0.2\linewidth}
\centering
    \resizebox{1.0\linewidth}{!}{
    \begin{tikzpicture}
    \begin{axis}[width=\linewidth,height=0.8\linewidth,
    ymode=log, log basis y=10,
    grid=both, scale only axis,
    xtick={0, 25, 50, 75, 100},
    ytick={1, 0.31622776601, 0.1, 0.05623413251, 0.031622776601, 0.0177827941, 0.01},
    xmin=0,
    xmax=100,
    axis background/.style={fill=gray!5},
    every axis x label/.style={at={(0.5, -0.1)},anchor=north, font=\sffamily\bfseries\small},
    every axis y label/.style={at={(0, 1)},anchor=north east, font=\sffamily\bfseries\small},
    ylabel=RMSE,
    xlabel=Step
    ]
    \addplot+[line width=2pt,mark=none,color=mycolor1]table[]{Figures/results/fandisk_step_vs_rmse_tonemapped_abs.tex};
    \end{axis}
    \end{tikzpicture}}
\end{minipage}

\begin{minipage}[b]{0.12\linewidth}\centering\textsf{\small Scene}\end{minipage}
\begin{minipage}[t]{0.2\linewidth}\centering\textsf{\small Cornell Box}\end{minipage}
\begin{minipage}[t]{0.2\linewidth}\centering\textsf{\small Sculpture}\end{minipage}
\begin{minipage}[t]{0.2\linewidth}\centering\textsf{\small Knots}\end{minipage}
\begin{minipage}[t]{0.2\linewidth}\centering\textsf{\small Fandisk}\end{minipage}

\begin{minipage}[b]{0.12\linewidth}\centering\textsf{\small RMSE}\end{minipage}
\begin{minipage}[t]{0.2\linewidth}\centering\small$0.0149$\end{minipage}
\begin{minipage}[t]{0.2\linewidth}\centering\small$ 0.0309$\end{minipage}
\begin{minipage}[t]{0.2\linewidth}\centering\small$0.0203$\end{minipage}
\begin{minipage}[t]{0.2\linewidth}\centering\small$0.0083$\end{minipage}

\begin{minipage}[b]{0.12\linewidth}\centering\textsf{\small Time (min)}\end{minipage}
\begin{minipage}[t]{0.2\linewidth}\centering\small$37.7$\end{minipage}
\begin{minipage}[t]{0.2\linewidth}\centering\small$170.4$\end{minipage}
\begin{minipage}[t]{0.2\linewidth}\centering\small$504.0$\end{minipage}
\begin{minipage}[t]{0.2\linewidth}\centering\small$412.3$\end{minipage}

\caption{Additional results. Our method performs well across diverse inputs, particularly those representing mostly smooth target images, including objects with glossy surfaces and complex topologies. Although it extracts only the exterior outlines of the geometry as domain boundaries at the start of optimization, the resulting handles tend to align well with color discontinuities caused by other geometric features and shadow boundaries. The computational cost is largely dictated by how effectively domain decomposition reduces the effective problem size. The Sculpture and Knots scenes are adapted from Zhao et al. \cite{Zhao:2018:InverseDiffusion}, with modifications.
}
\label{fig:results}
\end{figure*}

Our implementation is a multi-threaded C++/CUDA program. 
All the numbers reported in this paper were obtained using a desktop computer with two Intel Xeon Silver 4316 CPUs and an NVIDIA RTX A5000 GPU with 24 GB of GPU memory.
We use a modified version of the CPU renderer by \citet{Shirley:2024:raytracing}.  %
All the BEM-related operations are accelerated with a GPU; we generate renderer samples on the CPU first and send them as a batch to evaluate the BEM reconstructed solution and the Jacobian, which are then used to compute the update of parameters. We study our method mainly using the scene in \cref{fig:teaser} and also show additional results in \cref{fig:results}. We measure errors using RMSE against converged raster images of resolution $512^2$, after applying tone mapping. When we visualize diffusion curve handles, we visualize domain boundaries in black, the image border in gray, and handles within each domain using a consistent color unique to that domain.
All graphs use a logarithmic scale on the vertical axis.

\paragraph*{Progressive optimization}
Monte Carlo rendering into a raster image supports progressive rendering with errors appearing as noise. Our method also renders progressively, but the errors manifest as bias. \Cref{fig:progressive} shows several intermediate results for the scene in \cref{fig:teaser}. All other results shown in this paper correspond to the 100th step.

\paragraph*{Handle positions}
Compared to results without length regularization or endpoint snapping regularization, the full method produces more structured diffusion curves, with only a small increase in reconstruction error.
When the position update is disabled, the reconstructed image exhibits clear artifacts caused by the misalignment between handles and the target image’s color gradient (see \cref{fig:positions}).

\begin{figure}[h]
\centering
\hspace{-0.05\linewidth}%
\begin{minipage}[b]{0.35\linewidth}
\centering
\begin{tikzpicture}
\begin{axis}[
    width=\linewidth,
    height=0.58\linewidth,
    scale only axis,
    ymode=log, log basis y=10,
    grid=both,
    xtick={0, 25, 50, 75, 100},
    ytick={0.31622776601, 0.1, 0.031622776601, 0.01},
    xmin=0, xmax=100,
    axis background/.style={fill=gray!5},
    tick label style={font=\tiny},
    legend style={draw=none, font=\sffamily\bfseries\scriptsize, fill=none},
    legend cell align={left},
    legend pos=north east,
    every axis x label/.style={at={(0.5, -0.1)},anchor=north, font=\sffamily\bfseries\scriptsize},
    every axis y label/.style={at={(0, 1)},anchor=north east, font=\sffamily\bfseries\scriptsize},
]
\addplot+[line width=1.5pt,mark=none,color=mycolor1]table[]{Figures/results/cb2_data/nopruning_step_vs_rmse_tonemapped_abs.tex};
\addplot+[line width=1.5pt,mark=none,color=mycolor4]table[]{Figures/results/cb2_data/adamw_1e_1_step_vs_rmse_tonemapped_abs.tex};
\addplot+[line width=1.5pt,mark=none,color=mycolor2]table[]{Figures/results/cb2_data/adamw_1e_2_step_vs_rmse_tonemapped_abs.tex};
\addplot+[line width=1.5pt,mark=none,color=mycolor3]table[]{Figures/results/cb2_data/adamw_1e_3_step_vs_rmse_tonemapped_abs.tex};
\end{axis}
\end{tikzpicture}
\end{minipage}%
\hspace{0.15\linewidth}
\includegraphics[width=0.25\linewidth]{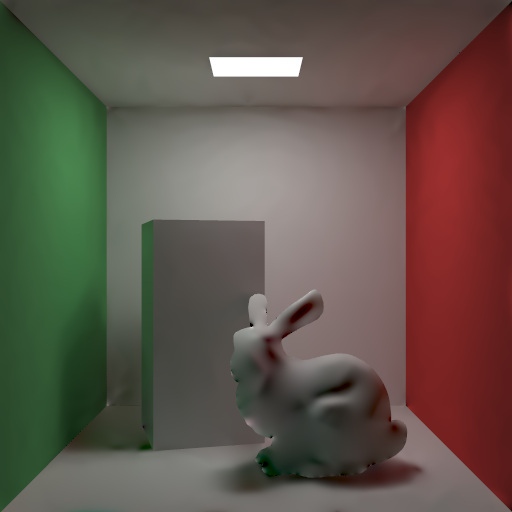}\\
\hspace{-0.05\linewidth}
\begin{minipage}[ht]{0.5\linewidth}\centering\textsf{\footnotesize RMSE / Step}\end{minipage}%
\begin{minipage}[ht]{0.25\linewidth}\centering\textsf{\footnotesize\color{mycolord1}Levenberg-\\\vspace{-1mm}Marquardt}\end{minipage}\\
\includegraphics[width=0.25\linewidth]{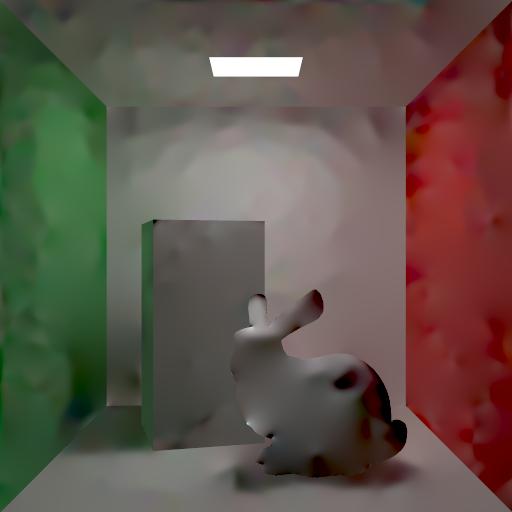}
\includegraphics[width=0.25\linewidth]{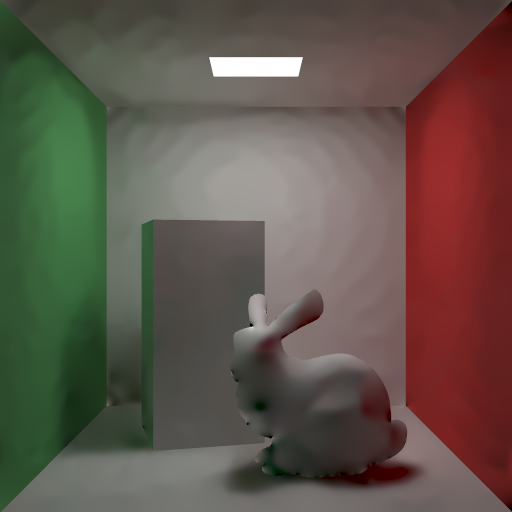}
\includegraphics[width=0.25\linewidth]{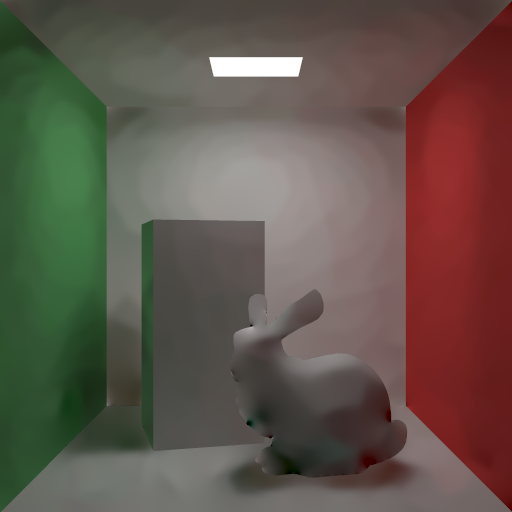}
\\
\begin{minipage}[ht]{0.25\linewidth}\centering\textsf{\footnotesize\color{mycolord3}AdamW\\\vspace{-1mm}(lr$=0.001$)}\end{minipage}
\begin{minipage}[ht]{0.25\linewidth}\centering\textsf{\footnotesize\color{mycolord2}AdamW\\\vspace{-1mm}(lr$=0.01$)}\end{minipage}
\begin{minipage}[ht]{0.25\linewidth}\centering\textsf{\footnotesize\color{mycolord4}AdamW\\\vspace{-1mm}(lr$=0.1$)}\end{minipage}
\\
\caption{Comparison of optimization methods. We compare Levenberg–Marquardt optimization with AdamW with three learning rates, scaled by powers of ten. Learning rates below 0.001 or above 0.1 result in slow convergence, while 0.01 performs best among the three. Nevertheless, even with this choice, AdamW converges more slowly than Levenberg–Marquardt.}
\label{fig:levenberg}
\end{figure}

\begin{figure}[h]
\centering
\begin{minipage}[b]{0.4\linewidth}
\centering
    \resizebox{1.0\linewidth}{!}{\begin{tikzpicture}
    \begin{axis}[width=1.2\linewidth,height=0.9\linewidth,
    ymode=log, log basis y=10,
    grid=both, scale only axis,
    xtick={0, 25, 50, 75, 100},
    ytick={1, 0.31622776601, 0.1, 0.05623413251, 0.031622776601, 0.0177827941, 0.01},
    xmin=0,
    xmax=100,
    axis background/.style={fill=gray!5},
    every axis x label/.style={at={(0.5, -0.1)},anchor=north, font=\sffamily\bfseries\small},
    every axis y label/.style={at={(0, 1)},anchor=north east, font=\sffamily\bfseries\small},
    ylabel=RMSE,
    xlabel=Step,
    legend style={draw=none, font=\sffamily\bfseries, fill=none},
    legend cell align={left},
    legend pos=north east,
    ]
    \addplot+[line width=2pt,mark=none,color=mycolor1]table[]{Figures/results/cb2_data/baseline_step_vs_rmse_tonemapped_abs.tex};
    \addlegendentry{\color{mycolor1}$\varepsilon=10^{-2}$}
    \addplot+[line width=2pt,mark=none,color=mycolor2]table[]{Figures/results/cb2_data/smallersmoothing_step_vs_rmse_tonemapped_abs.tex};
    \addlegendentry{\color{mycolor2}$\varepsilon=10^{-4}$}
    \end{axis}
    \end{tikzpicture}}\\
\end{minipage}%
\begin{tikzpicture}
\node[inner sep=0pt] (img1) at (0,0)
  {\includegraphics[width=0.28\linewidth]{Figures/results/cb2/cb2_baseline_100.png}};
\node[anchor=north west] at (img1.north west) {\footnotesize\textsf{\textcolor{white}{$10^{-2}$}}};
\end{tikzpicture}
\begin{tikzpicture}
\node[inner sep=0pt] (img2) at (0,0)
  {\includegraphics[width=0.28\linewidth]{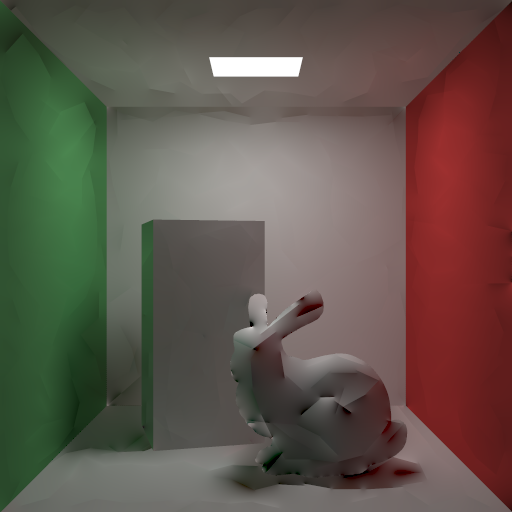}};
\node[anchor=north west] at (img2.north west) {\footnotesize\textsf{\textcolor{white}{$10^{-4}$}}};
\end{tikzpicture}
\caption{Discontinuity smoothing. We use a relatively large smoothing parameter of $\varepsilon = 10^{-2}$ (1\% of the image width) in all experiments. Reducing $\varepsilon$ to $10^{-4}$ increases the quantitative reconstruction error and introduces perceptual artifacts in the form of sharp color discontinuities at undesired locations.}
\label{fig:smoothing}
\end{figure}

\paragraph*{Levenberg–Marquardt}
We compare our optimization approach, based on the Levenberg–Marquardt method, against AdamW\linebreak\cite{AdamW2019} with three different learning rates  (\cref{fig:levenberg}).
With AdamW, too small or too large a learning rate leads to slow convergence, and even the best-performing setting still converges more slowly than Levenberg–Marquardt.
We believe this is because Levenberg–Marquardt is able to handle parameters with varying scales more effectively.
In addition to the scale differences between weights and position parameters, further scale variation arises across different domains.
A notable practical benefit is that Levenberg–Marquardt performs well without requiring hyperparameter tuning.

\paragraph*{Discontinuity smoothing}
Using a relatively large discontinuity smoothing parameter leads to more efficient optimization, consistent with expectations for problems involving discontinuities. In particular, when representing smooth images, employing handles with smoothed discontinuities tends to be more effective (\cref{fig:smoothing}).

\begin{figure}[h]
\centering
\hspace{-0.15\linewidth}%
\begin{minipage}[b]{0.35\linewidth}
\centering
\begin{tikzpicture}
\begin{axis}[
    width=\linewidth,
    height=0.7\linewidth,
    scale only axis,
    ymode=log, log basis y=10,
    grid=both,
    xtick={0, 25, 50, 75, 100},
    ytick={1, 0.31622776601, 0.1, 0.05623413251, 0.031622776601, 0.0177827941, 0.01},
    xmin=0, xmax=100,
    axis background/.style={fill=gray!5},
    ylabel=RMSE,
    xlabel=Step,
    tick label style={font=\tiny},
    legend style={draw=none, font=\sffamily\bfseries\scriptsize, fill=none},
    legend cell align={left},
    legend pos=north east,
    every axis x label/.style={at={(0.5, -0.1)},anchor=north, font=\sffamily\bfseries\scriptsize},
    every axis y label/.style={at={(0, 1)},anchor=north east, font=\sffamily\bfseries\scriptsize},
]
\addplot+[line width=1.5pt,mark=none,color=mycolor3] table[]{Figures/results/cb2_data/sparsity_100x_step_vs_rmse_tonemapped_abs.tex};
\addlegendentry{\color{mycolor3}{100x}}
\addplot+[line width=1.5pt,mark=none,color=mycolor2] table[]{Figures/results/cb2_data/sparsity_10x_step_vs_rmse_tonemapped_abs.tex};
\addlegendentry{\color{mycolor2}{10x}}
\addplot+[line width=1.5pt,mark=none,color=mycolor4] table[]{Figures/results/cb2_data/sparsity_1x_step_vs_rmse_tonemapped_abs.tex};
\addlegendentry{\color{mycolor4}{1x}}
\addplot+[line width=1.5pt,mark=none,color=mycolor1] table[]{Figures/results/cb2_data/baseline_step_vs_rmse_tonemapped_abs.tex};
\addlegendentry{\color{mycolor1}{0x}}
\end{axis}
\end{tikzpicture}
\end{minipage}%
\hspace{0.1\linewidth}%
\begin{minipage}[b]{0.35\linewidth}
\centering
\begin{tikzpicture}
\begin{axis}[
    width=\linewidth,
    height=0.7\linewidth,
    scale only axis,
    ymode=log,
    log ticks with fixed point,
    grid=both,
    xtick={0, 25, 50, 75, 100},
    ytick={100, 200, 400},
    xmin=0, xmax=100,
    ymin=100, 
    axis background/.style={fill=gray!5},
    ylabel=\#Handles,
    xlabel=Step,
    tick label style={font=\tiny},
    legend style={draw=none, font=\sffamily\bfseries\scriptsize, fill=none},
    legend cell align={left},
    legend pos=north east,
    every axis x label/.style={at={(0.5, -0.1)},anchor=north, font=\sffamily\bfseries\scriptsize},
    every axis y label/.style={at={(0, 1)},anchor=north east, font=\sffamily\bfseries\scriptsize},
]
\addplot+[line width=1.5pt,mark=none,color=mycolor1] table[]{Figures/results/cb2_data/baseline_step_vs_num_handles.tex};
\addplot+[line width=1.5pt,mark=none,color=mycolor4] table[]{Figures/results/cb2_data/sparsity_1x_step_vs_num_handles.tex};
\addplot+[line width=1.5pt,mark=none,color=mycolor2] table[]{Figures/results/cb2_data/sparsity_10x_step_vs_num_handles.tex};
\addplot+[line width=1.5pt,mark=none,color=mycolor3] table[]{Figures/results/cb2_data/sparsity_100x_step_vs_num_handles.tex};
\end{axis}
\end{tikzpicture}
\end{minipage}
\includegraphics[width=0.245\linewidth]{Figures/results/cb2/cb2_baseline_100.png}%
\includegraphics[width=0.245\linewidth]{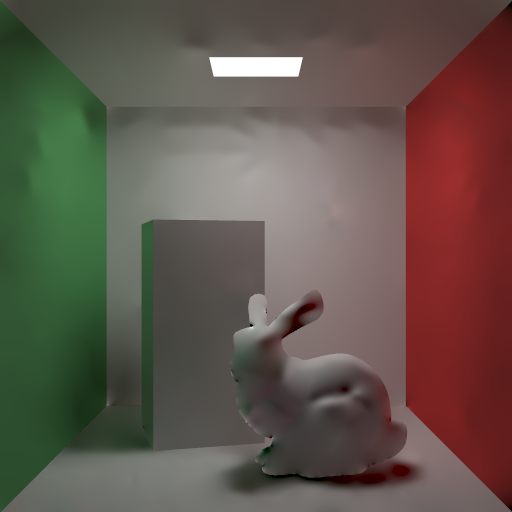}%
\includegraphics[width=0.245\linewidth]{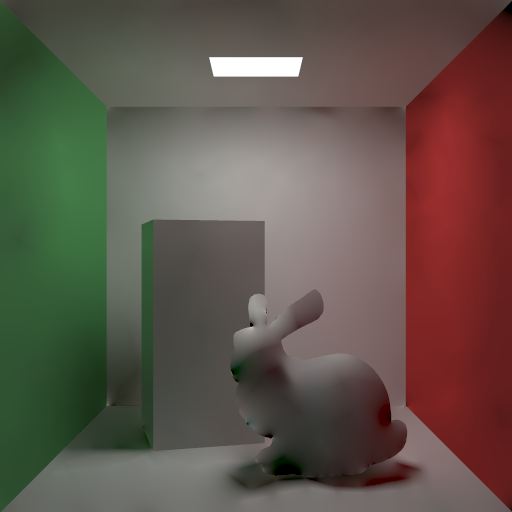}%
\includegraphics[width=0.245\linewidth]{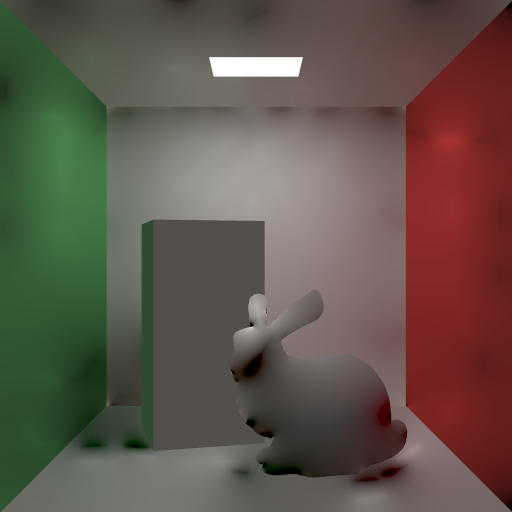}\\%
\includesvg[width=0.245\linewidth]{Figures/results/cb2/cb2_baseline_100.svg}%
\includesvg[width=0.245\linewidth]{Figures/results/cb2/cb2_sparsity_1x_100.svg}%
\includesvg[width=0.245\linewidth]{Figures/results/cb2/cb2_sparsity_10x_100.svg}%
\includesvg[width=0.245\linewidth]{Figures/results/cb2/cb2_sparsity_100x_100.svg}\\%
\begin{minipage}[b]{0.15\linewidth}\textsf{\small Sparsity:}\end{minipage}%
\begin{minipage}[b]{0.1\linewidth}\centering\textsf{\small {\color{mycolord1}0x}}\end{minipage}%
\begin{minipage}[b]{0.25\linewidth}\centering\textsf{\small {\color{mycolord4}1x}}\end{minipage}%
\begin{minipage}[b]{0.25\linewidth}\centering\textsf{\small {\color{mycolord2}10x}}\end{minipage}%
\begin{minipage}[b]{0.25\linewidth}\centering\textsf{\small {\color{mycolord3}100x}}\end{minipage}\\%
\begin{minipage}[b]{0.15\linewidth}\textsf{\small \#Handles:}\end{minipage}%
\begin{minipage}[b]{0.1\linewidth}\centering\textsf{\small $499$}\end{minipage}%
\begin{minipage}[b]{0.25\linewidth}\centering\textsf{\small $464$}\end{minipage}%
\begin{minipage}[b]{0.25\linewidth}\centering\textsf{\small $399$}\end{minipage}%
\begin{minipage}[b]{0.25\linewidth}\centering\textsf{\small $144$}\end{minipage}\\%
\begin{minipage}[b]{0.15\linewidth}\textsf{\small RMSE:}\end{minipage}%
\begin{minipage}[b]{0.1\linewidth}\centering{\small $0.0138$}\end{minipage}%
\begin{minipage}[b]{0.25\linewidth}\centering{\small $0.0163$}\end{minipage}%
\begin{minipage}[b]{0.25\linewidth}\centering{\small $0.0184$}\end{minipage}%
\begin{minipage}[b]{0.25\linewidth}\centering{\small $0.0222$}\end{minipage}%
\caption{Sparsity pruning. Results with different sparsity regularization constants $\lambda_\weight$. All optimizations start with $500$ handles, and we report the final number of handles and RMSE after 100 steps. Larger values of $\lambda_\weight$ produce sparser handle selections but also lead to higher errors, both quantitatively and qualitatively, manifesting as sharp color discontinuities and overly blurred regions.}
\label{fig:sparsity}
\end{figure}

\paragraph*{Sparsity pruning}
Our method allows users to optionally control handle sparsity by adjusting the sparsity regularization strength $\lambda_\weight$. We prune handles once the norms of the associated weights fall below a fixed threshold during the optimization. \cref{fig:sparsity} shows results without regularization (default) alongside those with increasing $\lambda_\weight$.
As expected, handle sparsity and reconstruction quality exhibit a tradeoff: higher regularization produces sparser solutions at the cost of increased reconstruction error. For smooth scenes, however, moderate increases in $\lambda_\weight$ can still yield good reconstructions, unless the regularization is too strong. To avoid the influence of errors introduced by sparsity regularization, we disabled it for all other experiments in this paper.

\begin{figure*}
\centering
\begin{minipage}[b]{0.12\linewidth}\centering\textsf{\small Reconstructed\\\;}\end{minipage}
\includegraphics[width=0.2\linewidth]{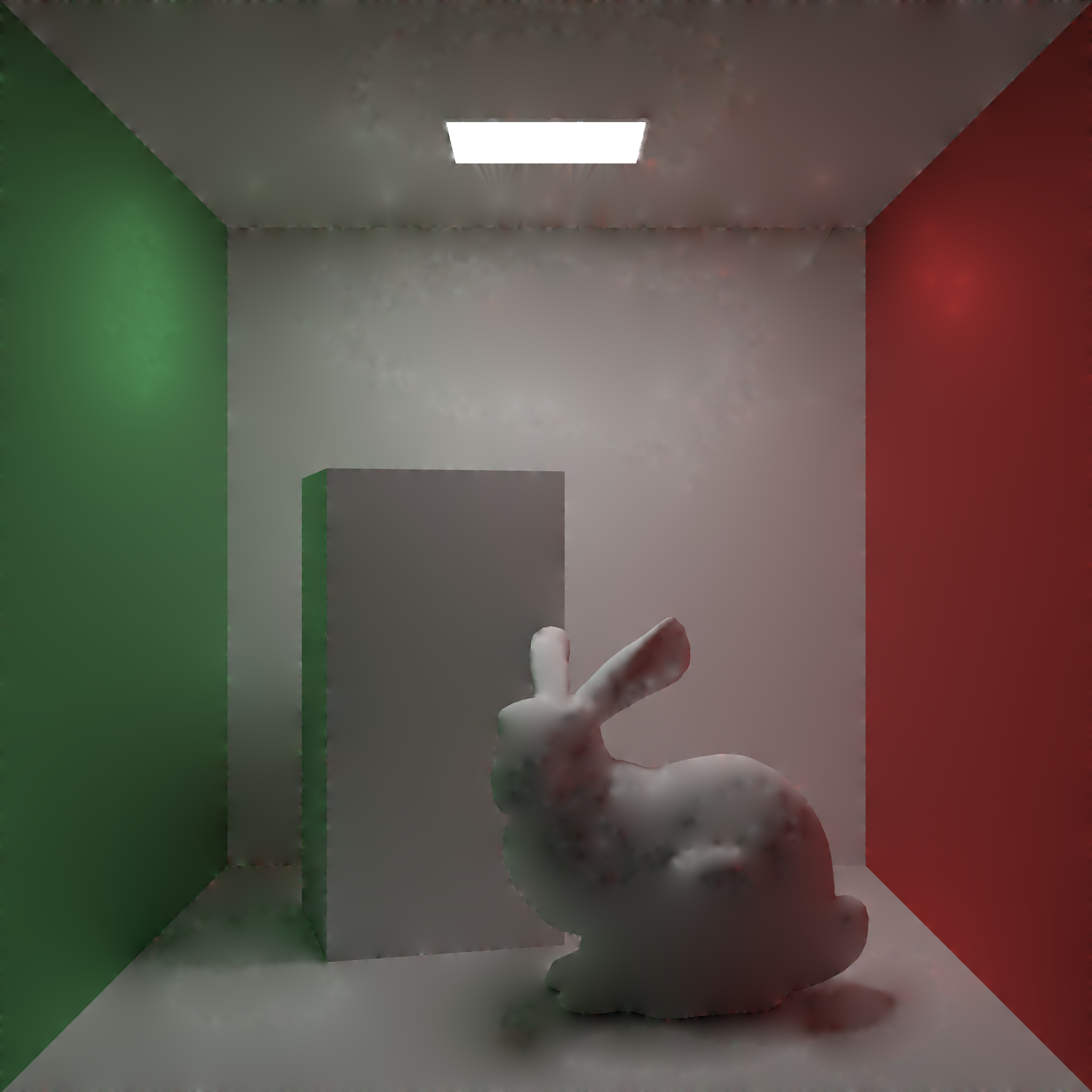}
\includegraphics[width=0.2\linewidth]{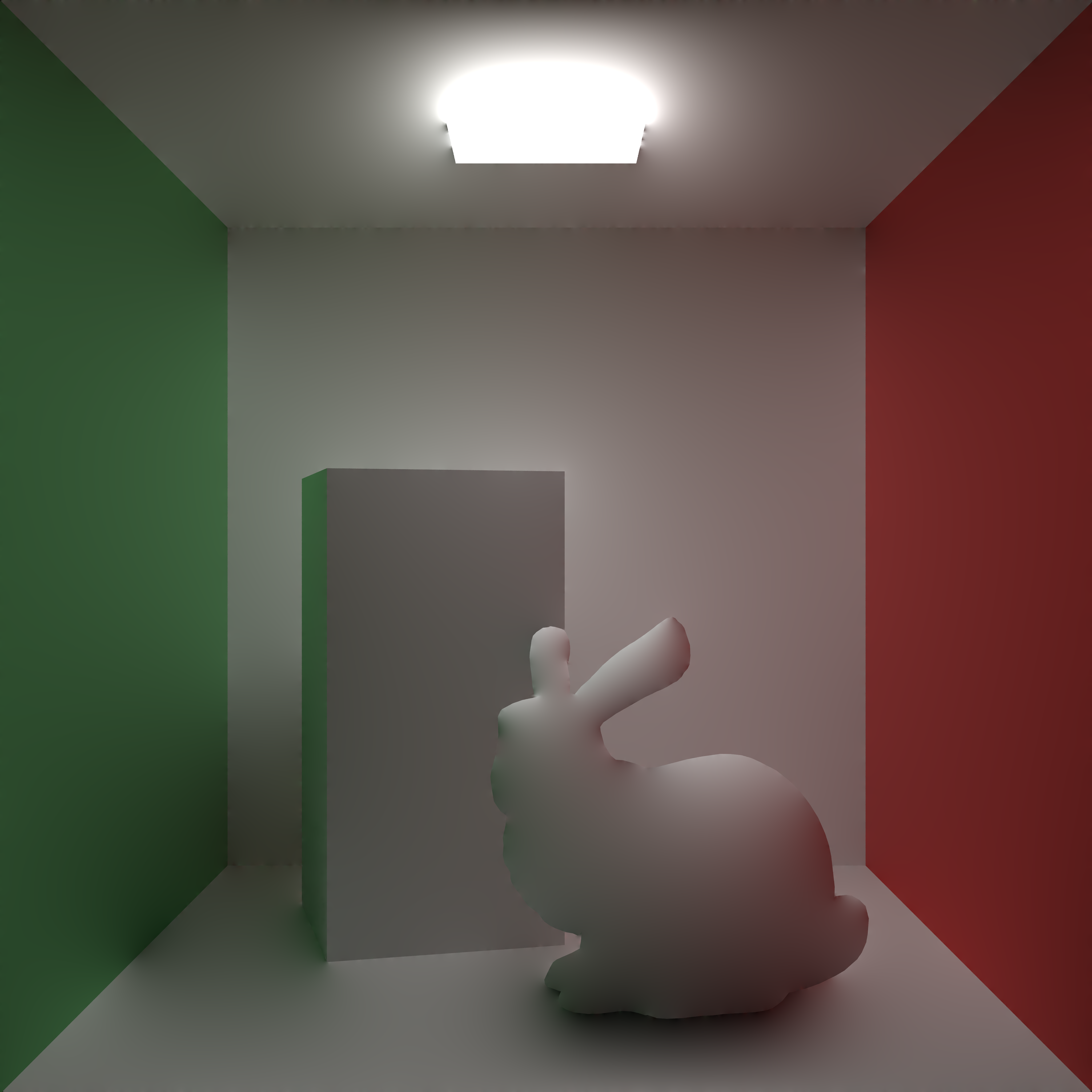}
\includegraphics[width=0.2\linewidth]{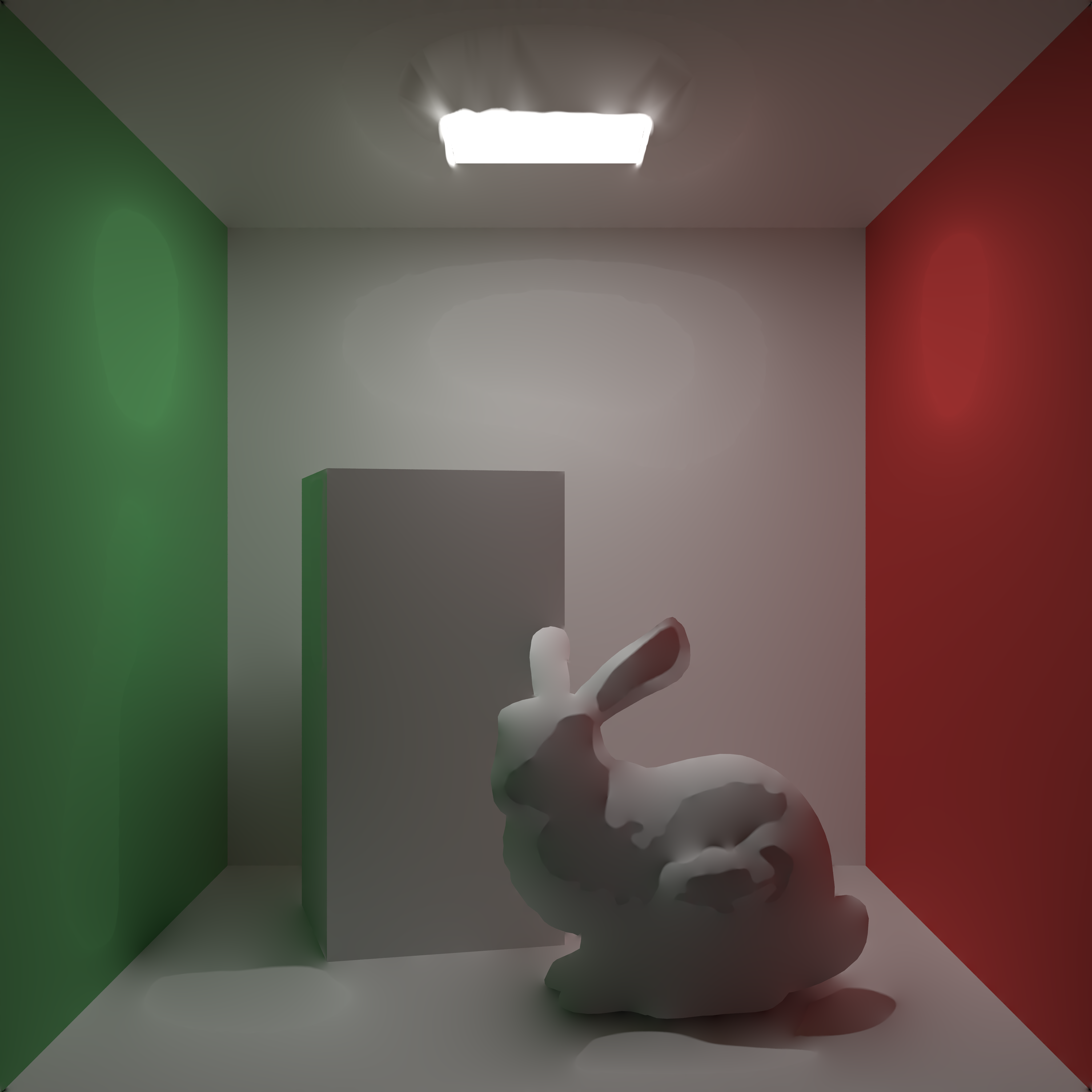}
\includegraphics[width=0.2\linewidth]{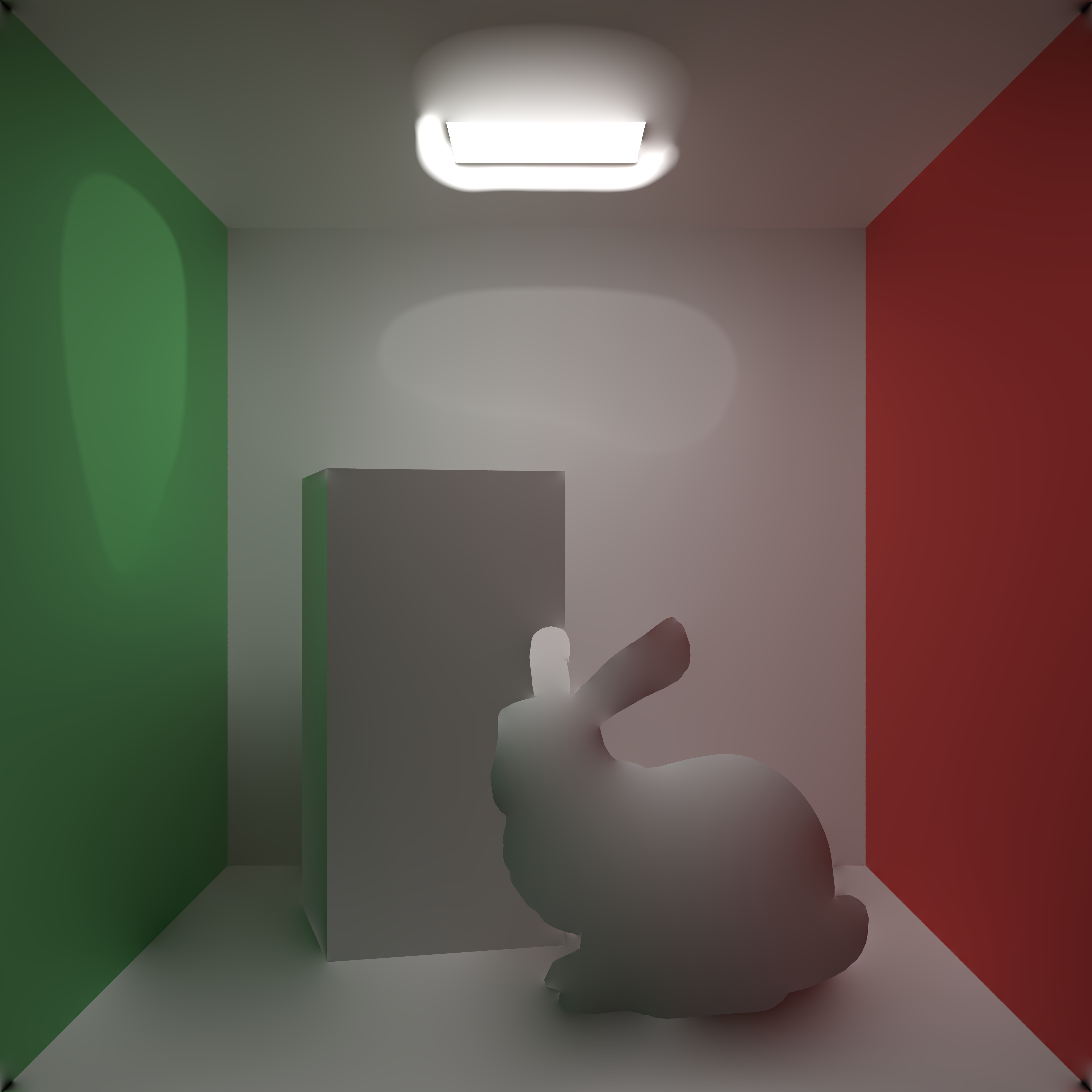}\\

\begin{minipage}[b]{0.12\linewidth}\centering\textsf{\small Diffusion\\Curves\\\;}\end{minipage}
\includesvg[width=0.2\linewidth]{Figures/fem/cbox2_4096.svg}
\includesvg[width=0.2\linewidth]{Figures/fem/cbox2_1024.svg}
\includesvg[width=0.2\linewidth]{Figures/fem/cbox2_256.svg}
\includesvg[width=0.2\linewidth]{Figures/fem/cbox2_64.svg}\\

\begin{minipage}[b]{0.12\linewidth}\centering\textsf{\small Resolution / Spp}\end{minipage}
\begin{minipage}[t]{0.2\linewidth}\centering\textsf{\small $4096^2\;/\;77$}\end{minipage}
\begin{minipage}[t]{0.2\linewidth}\centering\textsf{\small $1024^2\;/\;1221$}\end{minipage}
\begin{minipage}[t]{0.2\linewidth}\centering\textsf{\small $256^2\;/\;19532$}\end{minipage}
\begin{minipage}[t]{0.2\linewidth}\centering\textsf{\small $64^2\;/\;312500$}\end{minipage}\\

\begin{minipage}[b]{0.12\linewidth}\centering\textsf{\small RMSE}\end{minipage}
\begin{minipage}[t]{0.2\linewidth}\centering\small$0.0214$\end{minipage}
\begin{minipage}[t]{0.2\linewidth}\centering\small$0.1169$\end{minipage}
\begin{minipage}[t]{0.2\linewidth}\centering\small$0.0436$\end{minipage}
\begin{minipage}[t]{0.2\linewidth}\centering\small$0.0830$\end{minipage}

\caption{FEM-based optimization of \citet{Zhao:2018:InverseDiffusion} can easily fail in the presence of noise in the input image.
We run their program on the scene in \cref{fig:teaser} and visualize the reconstructed image with the extracted diffusion curves: black curves indicate the input contours, and dark-yellow curves are extracted by their method.
To avoid runtime errors due to issues such as self-intersecting geometry contours, we manually cleaned up the automatically extracted geometry contours and used them as an input to their program.
We use approximately the same number of samples as with the 100 iterations of our algorithm to render the input raster images.
While we maintain the same total number of samples across different tests, we vary the image resolution, resulting in different samples per pixel (spp). At low spp, the input images exhibit significant noise.
The result at a $4096^2$ resolution yields the lowest reconstruction error among the tested resolutions, although the diffusion curves are geometrically noisy and the reconstructed image contains visually unpleasant noise due to their local diffusion curve color assignment procedure. For the result at resolution $1024^2$, the internal triangulation algorithm failed to satisfy a vertex count condition, triggering a fallback mechanism that discards certain handles. As resolution decreases further, the curves become smoother and less noisy, but increasingly fail to capture the fine details in the input. Across all resolutions, we consistently observe color leaking around the ceiling light.
}
\label{fig:fem}
\end{figure*}

\paragraph* {Robustness}
Our method performs well even with noisy renderer samples and incomplete domain boundaries, as illustrated in the Sculpture scene (\cref{fig:results}), where domain boundaries are neither closed nor free of self-intersections. In contrast, the approach of \citet{Zhao:2018:InverseDiffusion} requires clean contours for successful triangulation of subdomains and assumes a noise-free input image; it can easily fail when noise is present.
When using a Monte Carlo renderer with a limited ray budget, one potential workaround for the noise in the input image is to render a low-resolution image with a high sample count per pixel. However, as shown in \cref{fig:fem}, finding an optimal resolution that balances noise reduction with sufficient detail can be challenging. For our experiments, we used their original code with minimal modifications for compilation and were able to reproduce their results on the original high-resolution ($1024^2$), noise-free inputs. Under a limited ray budget, however, their method degrades significantly due to noise.
This sensitivity to Monte Carlo noise was not previously demonstrated in diffusion curve vectorization works.

\section{Discussion and Future Work}

\begin{figure}[t]
\centering
\includegraphics[width=0.32\linewidth]{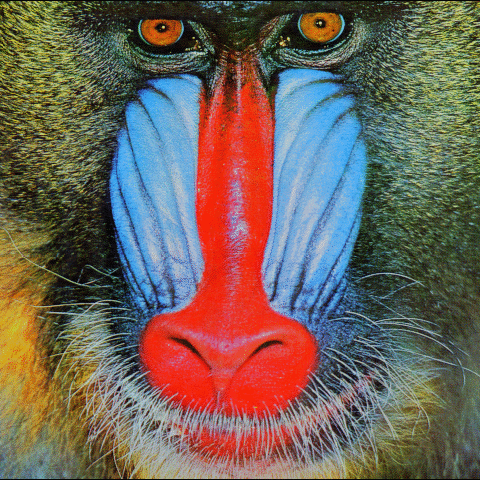}
\includegraphics[width=0.32\linewidth]{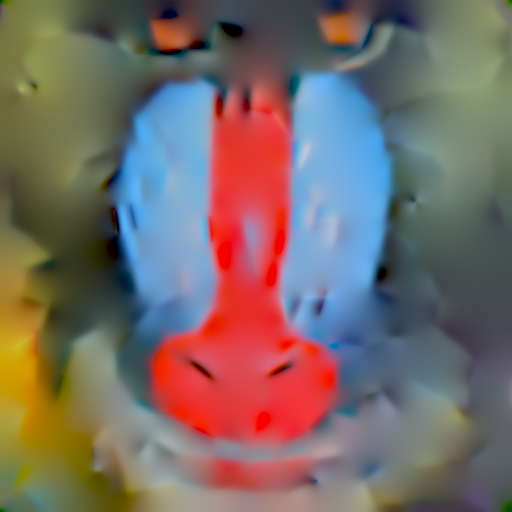}
\includesvg[width=0.32\linewidth]{Figures/results/baboon_100.svg}
\caption{Target image with high detail. While our method can be applied to input images with rich detail, it struggles to capture fine structures in high-frequency regions due to the limited number of handles we can have in our optimization. In such cases, our method primarily reproduces the overall color patterns, resulting in a blurred appearance due to the limited expressiveness of the finite number of handles.}
\label{fig:limitations}
\end{figure}

\paragraph*{Optimization}
We use the Levenberg-Marquardt method without convergence guarantees, assuming sufficiently accurate matrix estimates. To our knowledge, the only similar work in graphics is by \citet{Nicolet:2024:InverseRendering}, which applies a quasi-Newton method under similar assumptions. In deep learning, second-order methods for stochastic settings have gained attention~\cite{Mokhtari:2020:Survey, Guo:2023:Survey, Tian:2023:Survey, Bellavia:2018:LM}, though their stochasticity typically stems from data subsampling. By contrast, our setting adds noise per data point due to the noise of the Monte Carlo renderer. Designing optimization methods robust to this compounded noise is an open challenge.
Additionally, we currently approximate the loss Hessian using first-order gradients, but incorporating the full Hessian, similar to \citet{Eppler01022006}, could further improve accuracy and convergence.

\paragraph*{Efficiency}
In our BEM implementation, we parallelize only the naïve algorithm on the GPU to accelerate computation. Although established acceleration techniques for BEM, such as the fast multipole method and hierarchical matrices~\cite{Martinsson2019:fast, Chen:2025:Lightning}, are in principle applicable, their practical benefit remains unclear in our context due to the limited problem size.
These acceleration techniques might prove beneficial if our method were applied to target images with very high detail, requiring a large number of handles (\cref{fig:limitations}).
In our Monte Carlo optimization framework, we did not investigate importance sampling strategies, and especially treated the Monte Carlo renderer as a black box. A promising direction is to specially tune Monte Carlo sampling so that it accelerates the optimization process (e.g., stochastic gradient descent for training neural nets~\cite{Salaun:2025:GradientIS}). %

\paragraph*{Biharmonic diffusion curves}
As we can observe in some of the results, the diffusion curve vector image based on the Laplace equation is known to have some sharp color or color gradient discontinuities around handles~\cite{Orzan:Diffusion:2008}, and an additional blur step is required to smooth such discontinuities.
To address this problem, \citet{Finch:Freeform:2011} proposed an extension of the diffusion curves, 
which adopts the biharmonic equation instead of the Laplace equation. 
Forward BEM solvers for biharmonic diffusion curves are available~\cite{Ilbery:Biharmonic:2013, Xie:2014:Hierarchical}, but it is not yet clear how to generalize our differential solver framework to the biharmonic equation with some desired constraints on boundary conditions.

\paragraph*{Post-processing}
We focused on the problem of obtaining diffusion curve handles from a target image represented through noisy Monte Carlo samples. Once we obtain line segments with snapped endpoints, one could consider connecting line segments to produce polylines and even converting them to B\'ezier curves using the Potrace algorithm~\cite{selinger2003potrace}, as is done by \citet{Orzan:Diffusion:2008} and \citet{Zhao:2018:InverseDiffusion}.
If the errors introduced by such conversions are significant, one could optimize the geometry and colors further with a modified version of our proposed algorithm. 
Our approach with BEM should be extensible to other handle representations, such as B\'ezier curves, by adopting appropriate quadrature schemes.

\paragraph*{Downstream Applications}
Once the optimization is complete, the diffusion curve handles, together with the subdomain boundaries, define an interface that compactly represents the original 3D geometry and its shading in a reduced 2D space. An interesting future direction would be to leverage these handles to guide the optimization of the underlying 3D scene, drawing on techniques from inverse rendering~\cite{Zhao2020:PBDR}.

\paragraph*{Applicability of Differential BEM}
We presented an efficient differential BEM formulation for optimizing double-sided boundaries (handles) with optimizable boundary values, without requiring a linear system whose degrees of freedom correspond to these boundaries.  
This formulation can naturally generalize to problems involving single-sided boundaries with optimizable boundary values, by predefining one of the two jump values in \cref{eq:handleterm} to be zero. In this case, the method is equivalent to optimizing the boundary density of an indirect BEM formulation~\cite{Clemens:BEM2013}, still allowing joint optimization of boundary geometry and boundary values without introducing a linear system over those degrees of freedom.  
When boundaries have predefined boundary conditions, however, a linear system with the corresponding degrees of freedom becomes unavoidable, as in our treatment of outer zero Neumann boundaries. If the geometry of such boundaries is also subject to optimization, the linear system must be updated throughout the optimization process.  
Within these constraints, our differential BEM formulation should be applicable to a broader class of problems beyond the diffusion curve application considered in this work.

\begin{acks}
We thank Shuang Zhao for providing the code and scene data from their work, and the anonymous reviewers for their thoughtful and insightful feedback. This project was initiated while the first author was an intern at Adobe Research. This research was partially supported by NSERC Discovery Grants (RGPIN-2021-02524 \& RGPIN-2020-03918), CFI-JELF (Grant 40132), and a grant from Autodesk. The first author was partially supported by the David R. Cheriton Graduate Scholarship. AI-based tools were used to assist with improving the clarity and flow of the manuscript; all ideas and code implementations are solely attributable to the authors.
\end{acks}

\bibliographystyle{ACM-Reference-Format}
\bibliography{main}

\end{document}